\documentclass[structabstract]{aa}
\usepackage{graphicx}
\usepackage{txfonts}
\usepackage{natbib}
\bibpunct{(}{)}{;}{a}{}{,}

\begin{document}
   \title{A new method for calculating the convergent point of a \\moving group}

   \author{P.A.B. Galli
          \inst{1}
          \and
          R. Teixeira\inst{1}
          \and
          C. Ducourant\inst{2}
          \and
          C. Bertout\inst{3}
          \and P. Benevides-Soares\inst{1}
          }

   \institute{Instituto de Astronomia, Geof\'isica e Ci\^encias Atmosf\'ericas, Universidade de S\~ao Paulo, Rua do Mat\~ao, 1226 - Cidade Universit\'aria, 05508-900, S\~ao Paulo - SP, Brazil
         \and
             Observatoire Aquitain des Sciences de l 'Univers, CNRS-UMR 5804, BP 89, Floirac, France
          \and
          Institut d'Astrophysique, 98bis, Bd. Arago, 75014 Paris, France
             }

\date{Received / Accepted}

 \abstract{Convergent point (CP) search methods are important tools for studying the kinematic properties of open clusters and young associations whose members share the same spatial motion.}
 {We present a new CP search strategy based on proper motion data. We test the new algorithm on synthetic data and compare it with previous versions of the CP search method. As an illustration and validation of the new method we also present an application to the Hyades open cluster and a comparison with independent results.}
 {The new algorithm rests on the idea of representing the stellar proper motions by great circles over the celestial sphere and visualizing their intersections as the CP of the moving group. The new strategy combines a maximum-likelihood analysis for simultaneously determining the CP and selecting the most likely group members and a minimization procedure that returns a refined CP position and its uncertainties. The method allows one to correct for internal motions within the group and takes into account that the stars in the group lie at different distances.  }
 {Based on Monte Carlo simulations, we find that the new CP search method in many cases returns a more precise solution than its previous versions. The new method is able to find and eliminate more field stars in the sample and is not biased towards distant stars.  The CP solution for the Hyades open cluster is in excellent agreement with previous determinations. }
 {}


\maketitle


\section{Introduction}

Ever since their discovery, the existence of stellar groups with common space motion in the solar neighborhood has been an intriguing issue whose understanding is still far from complete. The origin and evolution of these comoving groups of stars, usually referred to simply as moving groups, is explained by different scenarios including cluster disruption, dynamical effects, and accretion events \citep{Eggen(1996),Dehnen(2000),Fux(2001),Navarro(2004)}.  Moving groups, which are observed as overdensities in the velocity space and exhibit a low internal velocity dispersion, typically a few km/s or less \citep{Mathieu1986}, allow study of the large-scale structure and dynamics of the Milky Way \citep{Antoja2008}. Because of perspective effects, the proper motions of comoving stars\footnote{In fact, the proper motions of a group of stars converge to a vertex either if their space motions are parallel or if they are expanding uniformly from a moving point. These two dynamical states are strictly equivalent as far as proper motions are concerned and radial velocity information is needed to distinguish between them \citep{Blaauw1964}.} appear to converge to a single point in the celestial sphere referred to as the \textit{convergent point} (hereafter CP) of the moving cluster. The CP is important not only for determining which stars are actual members of a moving cluster, but also for deriving individual kinematic distances of moving group members, provided that their radial velocities are known. This is very valuable when the trigonometric parallax from the ground is not accessible and \rm{\sc{Hipparcos}} parallaxes are not available \citep[for recent applications of this strategy see][]{Mamajek2005, Bertout2006}.

The first algorithms implementing a method for calculating the CP coordinates come from Charlier and Bohlin in 1916 \citep[see][]{Smart1968}. Each of them derived an equation independently using the position angle of stars to determine the CP position on the celestial sphere. However, the constant least-square coefficients involved in solving these equations were subject to measurement errors, thus leading to systematic errors in the derived CP coordinates. Later, \citet{Seares1945}, \citet{Petrie1949} and \citet{Roman1949} proposed different ways of correcting the Charlier and Bohlin equations. Their strategies represent, in a first approximation, different ways of using the position angle of stars to determine the CP solution.

Another approach was proposed by \citet{Brown1950}. He introduced a reference frame with the origin approximately at the center of the cluster and a fundamental plane defined by drawing a great circle through the origin in the direction of the average proper motion of the cluster. The proper motion of each star is then resolved into two components, one parallel to the reference plane and the other one perpendicular to the same plane, with the former expected to be much larger than the latter. The coordinates of the CP were then derived by applying the method of maximum likelihood.

Based on this method, \citet[][hereafter J71]{Jones1971} presented a  twofold algorithm dedicated to simultaneously  selecting the moving group members and calculating the CP position. This method was later improved and reformulated by \citet[][hereafter B99]{deBruijne1999} to make full use of the \textit{Hipparcos} data and allow for internal motions within the moving group.

There are many methods of finding the CP coordinates of moving groups. Several of them use, as J71 and B99 do, the observed proper motions of stars and differ in the details of the search strategy \citep[see for example][]{Makarov2001, Makarov2007a, Makarov2007b}. Other methods use different observed quantities, such as parallaxes and radial velocities \citep[]{Chen1997, Chereul1999, Hoogerwerf1999,Asiain1999}. The choice of which method to use depends essentially on the observational information that is available. The new CP search method that we propose in this article uses the observed proper motions to select the stars that belong to the moving group and determine the CP position. It builds on the works of J71 and B99, so we give in Sect.~2 some details on these methods that will be useful in presenting our own work. 

In the rest of this paper we present and test a new version of the CP search method. Section~3 describes our new algorithm while Sect.~4 deals with the construction of synthetic data from moving group simulations dedicated to testing and investigating the performance of our algorithm in comparison with previous ones. Section~5 presents an application of our method to the Hyades open cluster using \rm{\sc{Hipparcos}} data, and finally Sect.~6 summarizes the results of this work.


\section{The classic CP search method}

\subsection{Jones'  CP search method}

In J71, Jones proposed a maximum-likelihood method for simultaneously determining the CP position and the members of the moving group. To begin with, we consider a moving group of stars with known positions $(\alpha,\delta)$ and proper motions $(\mu_{\alpha}\cos\delta,\mu_{\delta})$, which are the two components of the proper motion vector $\mathbf{\mu}$. The tangential velocity $\mathbf{V_{t}}$ [km/s] is written in terms of the parallax $\pi$ [mas] of the star and its proper motion $\mathbf{\mu}$ [mas/yr] as

\begin{equation}\label{eq.1}
 \left|\mathbf{V_{t}}\right|=\frac{A\,\left|\mathbf{\mu}\right|\,}{\pi\,}{
}\, ,
\end{equation}
where $A=4.74047$ km yr/s is the ratio of one astronomical unit in km to the number of seconds in one Julian year. Furthermore, the radial velocity $\mathbf{V_{r}}$ [km/s] of a given star is given as a function of the stream space velocity $\mathbf{V}$ [km/s] by

\begin{equation}\label{eq.2}
 \left|\mathbf{V_{r}}\right|=\left|\mathbf{V}\right|\,\cos\lambda{
}\, ,
\end{equation}
where $\lambda$, the angular distance between the CP coordinates $(\alpha_{cp},\delta_{cp})$ and a given star of the moving group, is computed from

\begin{equation}\label{eq.3}
 \cos\lambda=\sin\delta\sin\delta_{cp}+\cos\delta\cos\delta_{cp}\cos(\alpha_{cp}-\alpha){
}\,.
\end{equation}
We recall that $\lambda$ is also the angle between $\mathbf{V}$ and $\mathbf{V_{r}}$ for parallel motions, which allows us to calculate the tangential and radial components of the space velocity. The two components of the proper motion $\mu_{\parallel}$, directed parallel to the great circle that joins the star and the CP, while $\mu_{\perp}$, directed perpendicular to the same great circle  (see Fig.~1), can be expressed as
\begin{equation}\label{eq.4}
\begin{array}{c}

\left(
  \begin{array}{r}
    \mu_{\parallel} \\
    \mu_{\perp} \\
  \end{array}
\right)
=
\left(
  \begin{array}{cc}
    \sin\theta & \cos\theta \\
    -\cos\theta & \sin\theta \\
  \end{array}
\right)
\left(
  \begin{array}{c}
    \mu_{\alpha}\cos\delta \\
    \mu_{\delta} \\
  \end{array}
\right)

\end{array}
\end{equation}
where the position angle $\theta$ of the CP is given by

\begin{equation}\label{eq.5}
 \tan\theta=\frac{\sin(\alpha_{cp}-\alpha)}{\cos\delta\tan\delta_{cp}-\sin\delta\cos(\alpha_{cp}-\alpha)}{
}\, .
\end{equation}

\begin{figure}[!htp]
\begin{center}
\includegraphics[width=0.40\textwidth]{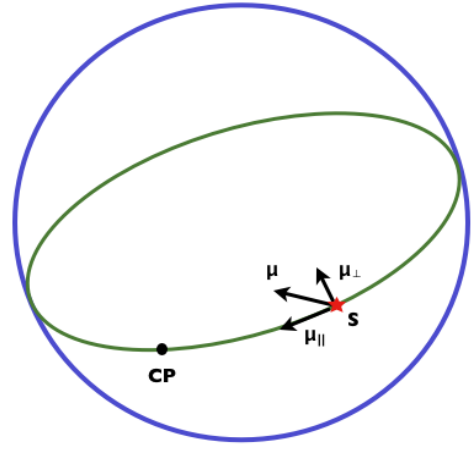}
\caption{Schematic representation of the proper motion components parallel and perpendicular to the great circle that contains the star S and the CP of the moving group.}
\end{center}
\label{fig1}
\end{figure}

When strict convergence occurs, i.e., when the proper motion vector is directed exactly towards the CP, we have $\mu_{\perp}=0$ and $\mu_{\parallel}=\mu$. In practice, strict convergence is not attained because of measurement errors and the internal velocity dispersion within the comoving group of stars.

The basic idea of Jones' method is to determine the CP of the moving group by comparing $\mu_{\perp}$ for each star with its expectation value of zero. The procedure for finding the CP of a moving group with $N$ stellar candidates starts by overlaying the sky with a grid. Each grid point is a CP candidate whose coordinates are denoted by $(\alpha_{cp},\delta_{cp})$. At each grid point one computes the $X^{2}$ value
\begin{equation}\label{eq.6}
X^{2}=$$\displaystyle\sum^{N}_{j=1}$$t^{2}_{\perp_{j}}{
 }\,
\end{equation}
where $t_{\perp}=\mu_{\perp}/ \sigma_{\perp}$ is the error-weighted value of $\mu_{\perp}$ for each star in the group. Assuming that $t_{\perp}$ is normally distributed, $X^{2}$ is distributed as $\chi^{2}$ with $N-2$ degrees of freedom. Minimizing $X^{2}$ is equivalent to maximizing the likelihood function of the computed $t_{\perp}$ values. Thus, the most likely CP is the grid point with the lowest $X^{2}$ value. However, the lowest $X^{2}$ could still occur by chance, so one should evaluate the probability $\epsilon$ that $X^{2}$ is higher than its observed value. It is given by
\begin{equation} \label{eq.7}
\epsilon=\frac{1}{\Gamma[\frac{1}{2}(N-2)]}$$\displaystyle\int^{\infty}_{X^{2}}x^{\frac{1}{2}[N-2]-1}e^{-x}dx{
}\,.
\end{equation}
If the computed probability is too low, the star with the highest $\left| t_{\perp}\right|$ is rejected. The number of stars in the sample is corrected and one goes back to the first step. This procedure is repeated until $\epsilon$ has reached a preset value (to be discussed later). When this is done, the grid point in the last iteration is defined as the CP of the moving group, and all remaining stars in the sample are considered to be group members. Thus, determining the CP of a moving group is linked to the selection of its members. Before doing the analysis discussed previously, it is necessary to eliminate stars with low $t_{\perp}$ values, because these stars are not rejected by the method. The rejection criterion is defined as
\begin{equation}\label{eq.8}
t=\frac{\mu}{\sigma}=\frac{\sqrt{\mu_{\alpha}^{2}\cos^{2}\delta+\mu_{\delta}^{2}}}{\sqrt{\sigma^{2}_{\mu_{\alpha}\cos\delta}+\sigma^{2}_{\mu_{\delta}}}}\leq
t_{min}{
}\, ,
\end{equation}
where $t_{min}$ is the rejection threshold (see below).

Searching for the CP on a grid of trial points defined on the sky was made necessary because of  the limited computer power available in the early 70s. However, the grid-based search represents a robust method of determining the CP position of a moving group, because it avoids the sometimes complicated topology of the $X^{2}$ function. One disadvantage of this implementation is that it only returns the coordinates of the CP and gives no information about their errors.

\subsection{De Bruijne's  CP search method}

Unlike the Jones method, which does not consider any possible internal motions within the moving group, the ``refurbished" CP method developed by B99 also handles this and takes advantage of the much more powerful computing facilities available today. Jones's  version of the CP method is modified in three ways and extended to include the determination of individual membership probabilities for each moving group star as explained below.

First, the method is adapted to take advantage of the \rm{\sc{Hipparcos}} data regarding the propagation of errors. Second, the definition of $t_{\perp}$ is modified to include the velocity dispersion in the moving group. The selection of stars with $t_{\perp}=0$ does not allow one to identify all moving group members, since the proper motions of some are not directed exactly towards the CP because of their velocity dispersion. De Bruijne's method assumes that the velocity dispersion $\sigma_{v}$ [km/s] and the mean distance $d$ [pc] of the group are known in advance. Thus, the one-dimensional velocity dispersion $\sigma_{int}$ [mas/yr]  of the group, in proper motion units, is given by

\begin{equation}\label{eq.9}
\sigma_{int}=\frac{1000\sigma_{v}\,}{A\,d\,}{
}\, ,
\end{equation}
and the new definition of $t_{\perp}$ becomes
\begin{equation}\label{eq.10}
t_{\perp}=\frac{\mu_{\perp}}{\sqrt{\sigma_{\perp}^{2}+\sigma_{int}^{2}}}{
}\, .
\end{equation}
The definition of $t_{min}$ is also modified to account for $\sigma_{int}$. It is written
\begin{equation}\label{eq.11}
t_{min}=\frac{\mu}{\sqrt{\sigma_{\mu}^{2}+\sigma_{int}^{2}}}.
\end{equation}
Furthermore, the grid-based approach used by J71 is replaced by a direct minimization routine that returns the CP position, as well as its uncertainty. Finally, a membership probability $p$ is assigned to each moving group member and takes the velocity dispersion into account. It is given by
\begin{equation}\label{eq.12}
p=\exp\left[-\frac{1}{2}\left(\frac{\mu_{\perp}^{2}}{\sigma_{\perp}^{2}+\sigma_{int}^{2}}\right)\right].
\end{equation}

As discussed by B99, the rejection threshold for a given star depends not only on its individual membership probability, but also on the membership probability of all stars considered. Including the velocity dispersion term in the definition of $t_{\perp}$ tends to raise the individual membership probabilities, thus allowing the method to include stars that are not members of the moving group. Therefore, stars with very low membership probability, even if not rejected by the method, should be carefully analyzed. On the other hand, stars with high membership probability (i.e. low $t_{\perp}$) should also be verified, because the selection of moving group members is biased towards distant stars that generally have smaller proper motions, thus smaller $\mu_{\perp}$ components.

\section{A new CP search method}

The original implementation of the CP method considers the stellar proper motion vector in two separate steps: a directional decomposition of the proper motion into the components $\mu_{\parallel}$, $\mu_{\perp}$ followed by a minimization routine based on the amplitude of $\mu_{\perp}$. Our algorithm differs from previous ones by considering both direction and amplitude of the stellar proper motions at once in the minimization routine. As a result, we will see that the new method is less biased towards more distant stars. We also take into account the internal motions of group members by introducing an individual correction for each star that depends on its proper motion and accounts for the fact that stars within the group lie at different distances. Our method makes an initial guess of the CP position using a grid-based approach that is similar to J71. Once an approximate CP position is found, it is then refined by an analytical minimization routine to return a more accurate solution. This procedure not only guarantees the solution convergence but also returns both the CP coordinates and their uncertainties. We present below the details of our algorithm.

The apparent motion of cluster members over the celestial sphere takes place along the arcs of great circles. The idea of visualizing the proper motion of a star over a great circle was first introduced by Herschel in 1783 \citep[see][]{Trumpler1953} and has been used more recently by \citet{Abad2003} and \citet{Abad2005}. Let  $\mathbf{r}=(x,y,z)=(\cos\alpha\cos\delta,\sin\alpha\cos\delta,\sin\delta)$ denote the position of a star with coordinates $(\alpha,\delta)$ and proper motion $(\mu_\alpha, \mu_\delta)$, in the usual orthogonal equatorial coordinate system, in which the unit vectors $\hat\mathbf{x}$, $\hat\mathbf{y}$, $\hat\mathbf{z}$ point respectively to the vernal equinox, the point on the equator with $\alpha=90^{\circ}$, and the northern equatorial pole. We compute the angular velocity of the star as the time derivative of the position vector
\begin{equation}\label{eq.13}
\dot{\mathbf{r}}=
\mu_{\alpha}\cos\delta
\left(
      \begin{array}{c}
      -\sin\alpha \\
       \cos\alpha\\
       0\\
       \end{array}
\right)
+
\mu_{\delta}
\left(
      \begin{array}{c}
       -\cos\alpha\sin\delta\\
       -\sin\alpha\sin\delta\\
       \cos\delta\\
       \end{array}
\right).
\end{equation}
The great circle representing the motion of a given star is the intersection of the celestial sphere with the plane defined by $\mathbf{r}$ and $\dot{\mathbf{r}}$ (see Fig.~2). Whenever a group of stars with parallel motions exists, their great circles should intersect at two opposite points on the celestial sphere: the CP and its mirror point. We denote the mirror point of the CP in this work as the \textit{divergent point} (DP hereafter) of the moving group. Assuming that $(\alpha_{cp},\delta_{cp})$ are the coordinates of the CP, the coordinates of the DP are $(\alpha_{dp},\delta_{dp})=(\alpha_{cp}+180^{\circ},-\delta_{cp})$. From this visualization of the proper motions over the entire sphere one sees that a search for the group vertex will in fact return two geometrically equivalent solutions. The distinction between both solutions depends on the direction of the proper motion vectors.

The great circle of each moving group member contains the star, the CP (DP) and the direction of motion. It can be described by the vector $\mathbf{p}$ normal to the plane defined by $\mathbf{r}$ and $\dot{\mathbf{r}}$:
\begin{equation}\label{eq.14}
\mathbf{p}=\mathbf{r}\times\dot{\mathbf{r}}=
\mu_{\alpha}\cos\delta
\left(
      \begin{array}{c}
      -\cos\alpha\sin\delta \\
       -\sin\alpha\sin\delta\\
       \cos\delta\\
       \end{array}
\right)
-
\mu_{\delta}
\left(
      \begin{array}{c}
      -\sin\alpha \\
       \cos\alpha\\
       0\\
       \end{array}
\right).
\end{equation}
For each group member, the vector $\mathbf{p}$ defines the \textit{pole} of its great circle on the celestial sphere. The pole representation of each star is unique and includes its position plus its motion. Whenever a group of comoving stars exists, the poles of the great circles of its members are located in a common plane, and the intersection of this plane with the celestial sphere forms another great circle. The poles of this \textit{polar great circle} are the apex and antapex of the moving group, i.e., its CP and DP. In practice these two points are the intersections of the individual stellar great circles (see Fig.~2). They are calculated by forming the cross products of all poles of members between themselves, $\mathbf{p_{i}}\times\mathbf{p_{j}}$, when $i$ and $j$ are  group members. The distribution of intersection points between the polar great circles of a group of stars can highlight preferred directions of motion and may be used to search for moving group members \citep[see][]{Abad2003}.

\begin{figure}[!htp]
\begin{center}
\includegraphics[width=0.50\textwidth]{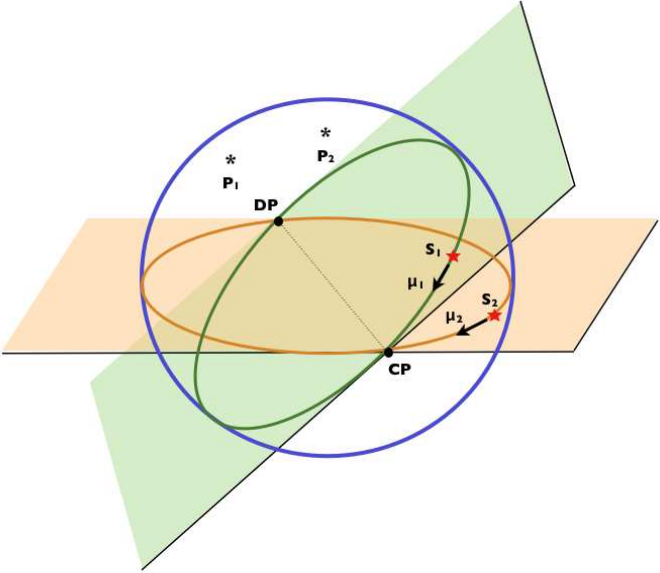}
\caption{Schematic representation of the motion over a great circle for two stars $S_{1}$ and $S_{2}$ of the moving group with proper motions $\mu_{1}$ and $\mu_{2}$. The poles of the corresponding great circles are given by $P_{1}$ and $P_{2}$. }
\end{center}
\label{fig2}
\end{figure}

\subsection{The new CP search algorithm}

The basic idea of our procedure is to determine the \textit{polar} great circle that best interpolates the individual great circle poles of the moving group stars. The poles of the interpolated great circle return the CP and DP solutions of the moving group. To derive a first approximation of the CP coordinates we follow the grid-based approach developed by J71, which is adapted to our strategy. For simplicity we describe the algorithm in the equatorial coordinate system, but it is independent of the chosen coordinate system.

Given a sample of $N$ stars with known positions $(\alpha,\delta)$, proper motions $(\mu_{\alpha}\cos\delta,\mu_{\delta})$, and corresponding errors, we perform the following steps:

\begin{enumerate}
\item Compute the pole $(\alpha^{p},\delta^{p})$ of the great circle for each star in the sample and the corresponding errors $(\sigma_{\alpha^{p}},\sigma_{\delta^{p}})$. The derived expressions are given in Appendix~A. 

\item Define a grid of trial CPs on the plane of the sky and number each grid point $(i=1, 2, 3, ..., N_{grid})$.

\item Start at grid point $i=1$ and assume that this point is the CP. The coordinates of this point are referred to as $(\alpha_{cp},\delta_{cp})$.

\item Calculate for each star the error-weighted value $s$ of the orthogonality error $\kappa$

\begin{equation}\label{eq.15}
s=\frac{\kappa}{\sigma_{\kappa}}{
}\,.
\end{equation}
The orthogonality error $\kappa$ is defined as
\begin{equation}\label{eq.16}
\kappa_{j}=\sin\delta_{cp}\sin\delta_{j}^{p}+\cos\delta_{cp}\cos\delta_{j}^{p}\cos(\alpha_{cp}-\alpha_{j}^{p}),
\end{equation}
where $j$ runs over all stars in the sample $(j=1,2,3,...,N)$. The derived expressions for $\kappa$ and $\sigma_{\kappa}$ are given in Appendix~B. We assume that $s$ is normally distributed with zero mean and unit variance (the validity of this hypothesis will be discussed in Sect.~4.5). The probability distribution for the star $j$ with a given $s$ is then

\begin{equation}\label{eq.17}
p_{j}=\frac{1}{\sqrt{2\pi}}\exp\left(-\frac{1}{2}s_{j}^{2}\right).
\end{equation}

\item At the given grid point compute $X^{2}$ as
\begin{equation}\label{eq.18}
X^{2}=$$\displaystyle\sum^{N}_{j=1}$$s^{2}_{j}{
 }\,.
\end{equation}

\item Determine $X^{2}$ at each grid point by repeating steps 3-5.

\item The total probability $P$ for the set of calculated values is given by
\begin{equation}\label{eq.19}
P=$$\displaystyle\prod^{N}_{j=1}p_{j}$$=\frac{1}{(2\pi)^{N/2}}e^{-\frac{1}{2}X^{2}}{
}\,,
\end{equation}
and defines the likelihood function. With $s$ distributed normally, $X^{2}$ is distributed as $\chi^{2}$ with $N-2$ degrees of freedom. Maximizing the likelihood function is equivalent to minimizing $X^{2}$.  Define the grid point with the lowest $X ^{2}$ as the most likely CP.

\item The lowest $X^{2}$ could still occur by chance rather than by a good fit between the observations and the model. As in J71, evaluate the probability $\epsilon$ that $X^{2}$ will exceed the observed value of $X^{2}$ by chance. It is given by
\begin{equation}\label{eq.20}
\epsilon=\frac{1}{\Gamma[\frac{1}{2}(N-2)]}$$\displaystyle\int^{\infty}_{X^{2}}x^{\frac{1}{2}[N-2]-1}e^{-x}dx{
}\,,
\end{equation}
where $\Gamma(x)$ denotes the Gamma function for $x>0$.

\item If $\epsilon < \epsilon_{min}$, that is to say, if the computed value of $X^{2}$ is unacceptably high, reject the star with the highest $|s|$ value, correct the number of stars in the sample $N\rightarrow N-1$, and go back to step 3. Otherwise continue to the next step.

\item When $\epsilon$ has reached an acceptable value (to be discussed in Sect.~4.4), choose the grid point considered in the last iteration as the maximum-likelihood CP and all non rejected stars in the sample are identified as moving group members.
\end{enumerate}

The algorithm presented above simultaneously selects moving group members and calculates the CP position. It is based only on position and proper motion data. As in J71, before starting the procedure, it is necessary to reject all stars with proper motion data that carry poor information because of measurement errors.

\subsection{Correction for internal velocity dispersion}

The new CP search algorithm described above selects those stars in the sample as moving group members whose orthogonality error $\kappa$ approaches zero. When $\kappa$ approaches its expected value of zero, the plane defined by the individual poles of stars and the one that passes through the CP and DP solutions are orthogonal. The poles of group members are located on the same plane when the stars in the group have parallel motions. However, perfect parallelism will not necessarily be achieved by some moving group members. Therefore, selecting only those stars with $s=0$ will not identify all group members. Consequently, a small amount of deviation of $s$ due to the velocity dispersion in the group should be permitted.

The procedure that we use to allow for internal motions within the moving group is similar to the one proposed by B99. The definition of $s$ is changed to

\begin{equation}\label{eq.21}
s=\frac{\kappa}{\sqrt{\sigma_{\kappa}^{2}+\Delta\kappa^{2}}}{
}\,,
\end{equation}
where $\Delta\kappa$ is an estimate of the one-dimensional velocity dispersion in the group that is translated into the proper motion scatter $(\Delta\mu_{\alpha},\Delta\mu_{\delta})$ of group stars. It is given by

\begin{equation}\label{eq.22}
\Delta\kappa=\sqrt{\left(\left|\frac{\partial\kappa}{\partial\mu_{\alpha}^{*}}\right|\Delta\mu_{\alpha}^{*}\right)^{2}+\left(\left|\frac{\partial\kappa}{\partial\mu_{\delta}}\right|\Delta\mu_{\delta}\right)^{2}},
\end{equation}
where $\mu_{\alpha}^{*}=\mu_{\alpha}\cos\delta$. The proper motion scatter in each component is estimated by using Eq.~(\ref{eq.9}) and assuming that the velocity dispersion and the mean distance to the group are known in advance. We constructed synthetic samples of moving groups (as discussed in Sect.~4) with different velocity dispersions and concluded that the scatter observed in the proper motion components is consistent with this estimate given by Eq.~(\ref{eq.9})  and similar in both components. One should note that the term $\Delta\kappa$ proposed in this work to take the internal motions within the group into account differs from the one introduced by B99, because the former is an individual correction applied to each star. Although average group parameters (velocity dispersion and distance) are used, the partial derivatives in Eq.~(\ref{eq.22}) depend on the stellar position and proper motion of each group member. Because we use the proper motions, we account for the various distances of group members. This point is particularly important when dealing with moving groups that occupy a large volume in space.

Following the procedure developed by B99, we then attribute a membership probability to each group member, defined as
\begin{equation}\label{eq.23}
p_{j}=\exp\left[-\frac{1}{2}\left(\frac{\kappa_{j}^{2}}{\sigma_{\kappa_{j}}^{2}+\Delta\kappa_{j}^{2}}\right)\right].
\end{equation}

\subsection{Refining the CP coordinates}

Once we have found an approximated CP position $(\alpha_{cp},\delta_{cp})$ of the moving group as explained above, we refine it by implementing a direct minimization routine in two dimensions that leads to the following non linear least square equations that must be solved. We have

\begin{equation}\label{eq.24}
\frac{\partial X^{2}}{\partial\alpha_{cp}}=0\hspace{0.1cm}\rightarrow\hspace{0.1cm}\sum^{N}_{j=1}s_{j}\frac{\partial s_{j}}{\partial\alpha_{cp}}=0
\end{equation}
\begin{equation}\label{eq.25}
\frac{\partial X^{2}}{\partial\delta_{cp}}=0\hspace{0.1cm}\rightarrow\hspace{0.1cm}\sum^{N}_{j=1}s_{j}\frac{\partial s_{j}}{\partial\delta_{cp}}=0.
\end{equation}
We follow the Levenberg-Marquardt method to solve these equations as described in \citet{Madsen2004} and \citet{Press1992}. The model is approximated at each iteration  by a first order Taylor series expansion, and we use the initial guess $(\alpha_{cp},\delta_{cp})$ of the CP position to obtain the final coordinates of the CP iteratively by successive approximations. This method returns the CP solution and its uncertainty in the form of a 2x2 covariance matrix. The stopping criterion for our routine is defined by the magnitude of the CP uncertainties in the covariance matrix since iterating to machine accuracy is generally unnecessary.

\bigskip
To simplify the following discussion, we denote the two CP search methods (CPSMs) discussed in Sect.~2 (J71 and B99) collectively by the name \emph{classic CP search method}, or classic CPSM, and the new method described in Sect.~3 by the name \emph{new CP search method}, or new CPSM. We emphasize that the definition of the $X^{2}$ function is not the same in the classic and new CPSMs. We used the appropriate definition of $X^{2}$ in the various tests of the classic and new CPSMs that we discuss below. Our minimization routine using the analytical derivatives $\partial X^{2}/\partial\alpha_{cp}$ and  $\partial X^{2}/\partial\delta_{cp}$ differs from the one used by B99.


\section{Monte Carlo simulations of moving groups}

CPSMs handle two tasks:  the search for the most likely moving group members and determination of the CP position. Detection of group members requires the moving group to be distinguished from the field population by its kinematic properties, whereas the position of the CP is influenced by several parameters. To investigate both abilities of the CP method, we applied the classic and new CPSMs to synthetic samples of moving groups with different configurations. In our simulations we focused mainly on the differences and advantages of both methods.

\subsection{Construction of synthetic datasets}

We construct our synthetic data using the Galactic coordinates system, which is more convenient, particularly when dealing with Galactic rotation. That is, we define a rectangular coordinate system in which the unit vectors $\mathbf{\hat{u}}$, $\mathbf{\hat{v}}$, $\mathbf{\hat{w}}$ point, respectively, to the Galactic center, to the direction of Galactic rotation, and to the Galactic north pole.  The Galactic coordinates $(l,b)$ of the stars are randomly drawn in the sky region occupied by the assumed moving group. We vary the Galactic longitude $l$ of the cluster center from $0^{\circ}$ to $360^{\circ}$ in steps of $60^{\circ}$ and assume a constant latitude of $b=0^{\circ}$. Our choice of $b$ is motivated by the fact that young clusters and star-forming regions lie very close to the Galactic plane. The individual distances are drawn from a Gaussian distribution where the mean distance depends on the cluster simulation. The stars are distributed in a distance range of 30~pc projected along the line of sight. We assume three different configurations of distance and vary the projected angular size of the group. Table~1 summarizes the characteristics of our synthetic samples.
\begin{table}[!htp]
\centering
\caption{Mean distance, field size, and number of moving group and field stars in the synthetic samples for each configuration (A,B,C) considered in this work.}
\begin{tabular}{ccccc}
\hline
\#&Distance (pc)&Field Size $l$ x $b$ (deg$^{2}$)&$N_{MG}$&$N_{field}$\\
\hline
A& 100& 20 x 20& 200&1000 \\
B& 200& 15 x 15&100&500 \\
C& 400& 10 x 10&50&250\\
\hline
\end{tabular}
\label{tab1}
\end{table}

The stellar velocity components $\mathbf{V}=(u,v,w)$ are randomly drawn from a sphere in velocity space with radius of 20km/s. The three velocity components and position are translated into proper motion $(\mu_{l}\cos b,\mu_{b})$ and radial velocity $V_{r}$ for each star, after which observational errors are added. Position and proper motion errors are also drawn from Gaussian distributions with means of 1~mas and 1~mas/yr, respectively. We then add to the streaming motion of the group three components representing the velocity dispersion, the reflex of the Solar motion with respect to the local standard of rest, and the Galactic rotation, in the following manner. Each of the dispersion velocity components $(\delta u,\delta v,\delta w)$ is independently drawn from a Gaussian distribution with an isotropic velocity dispersion of  $\sigma_{v}=1$ km/s. Our choice of $\sigma_{v}$ is motivated by the fact that young stellar groups often exhibit a low-velocity dispersion. For the solar motion we adopt $(U,V,W)_{\odot}=(10.00,5.23,7.17)$ km/s \citep[]{Dehnen1998} \footnote{Although a more recent value of the solar motion given by \citet[]{Schonrich(2010)} exists, we adopted the value of \citet[][]{Dehnen1998}, which has been widely used in the literature. The results and conclusions of this paper do not depend on the specific value used for the solar motion.} and $A=14.82$ km s$^{-1}$kpc$^{-1}$ and $B=-12.37$ kms$^{-1}$kpc$^{-1}$ for the Oort constants \citep[]{Feast1997}. We introduce Galactic rotation in the simulated  stellar proper motions by using the first-order formulae given in \citet[]{Green1985}. 

Our synthetic samples of stars include both the moving group stars and a field population. The density of stars in each field corresponds approximately to that of the \rm{\sc{Hipparcos}} catalog ($\sim$3 stars/deg$^{2}$). Field stars are located in the same field of view and have the same observational errors as moving groups stars. They are unrelated to the moving group and have random motions in amplitude and direction.

Each run consists of 100 Monte Carlo cluster simulations for each Galactic position and distance of the cluster. We follow the procedure adopted by B99 to correct for internal motions by taking the value for $\sigma_{v}$ that was actually used to construct our synthetic data samples. For both CPSMs, we assume the stop value of the probability $\epsilon$ defined in Eq.~\ref{eq.7} to be $\epsilon_{min}=0.954$, as suggested by B99, (we come back to this in Sect. 4.4). In the following, we investigate and compare both CPSMs using the synthetic data sets. We first investigate the ability of each CPSM to find group members and reject field stars (see Sect.~4.2), then we come to a more detailed analysis of the CP itself as a function of several parameters (see Sect.~4.3).

\subsection{Selection of cluster members and rejection of field stars}

The CP method selects moving group members based on their proper motions and reject field stars that are not in the same proper motion range. A distinction must be made between field stars at the same distance as the moving group and background stars. To begin with, we construct samples as described in Sect.~4.1 and consider the case where the field population is at the same average distance as the moving group. The results of these simulations are shown in Fig.~3. On the whole, both CPSMs exhibit a similar performance even if the fraction of rejected field stars by the new CPSM is higher in some specific cases . Our simulations show that at the shorter distances of Table~1 the fraction of cluster members that can be retrieved for both CPSMs is higher than $\sim 80\%$ and the contamination of field stars amounts to $\sim 20\%$ of the total number of field stars. At greater distances, the ability to find and eliminate field stars decreases for both CPSMs, as expected, since the quality of proper motions also diminishes.

We now consider the case where the field population consists of background stars located two times farther than the moving group. We perform the same simulations as before and present the results in Fig.~4. We find that the performance of the two methods differ; the new CPSM is likely to eliminate more background stars than the classic CPSM. The difference between both CPSMs is more evident at greater distances (e.g. at $d=$~400~pc) where the distributions of rejected field stars for each CPSM are clearly separated.  As already discussed by B99, the classic CPSM is biased toward distant stars. The definition of the $X^{2}$ function to be minimized in the classic CPSM considers only one directional component of the proper motion vector, which leads to a biased selection of stars with small $\mu_{\perp}$ components. However, not all stars with low $\mu_{\perp}$ are necessarily members, because stars at farther distances, such as background stars, generally have smaller proper motions (i.e. small $\mu_{\perp}$). On the other hand, the definition of the $X^{2}$ function using great circle poles in the new CPSM considers both direction and amplitude of the proper motion vector at the same time. By considering the \textit{full} proper motion vector instead of only one directional component (as in the classic CPSM) in the definition of the minimizing function $X^{2}$ leads to a membership selection that is not biased towards distant stars. 

\begin{figure}[!htp]
\begin{center}
\vspace{0.2cm}
\includegraphics[width=8cm]{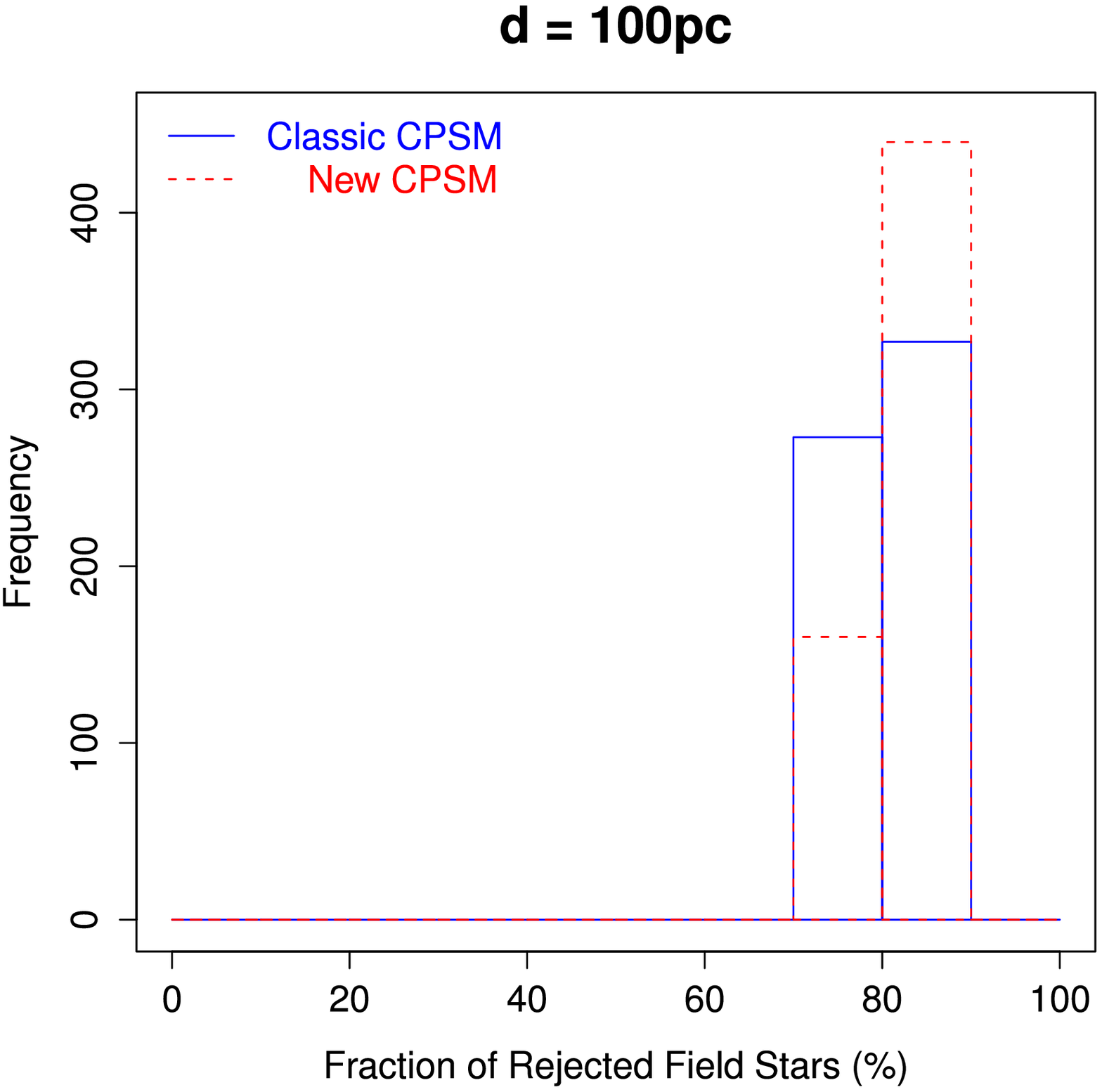}\vspace{0.4cm}
\includegraphics[width=8cm]{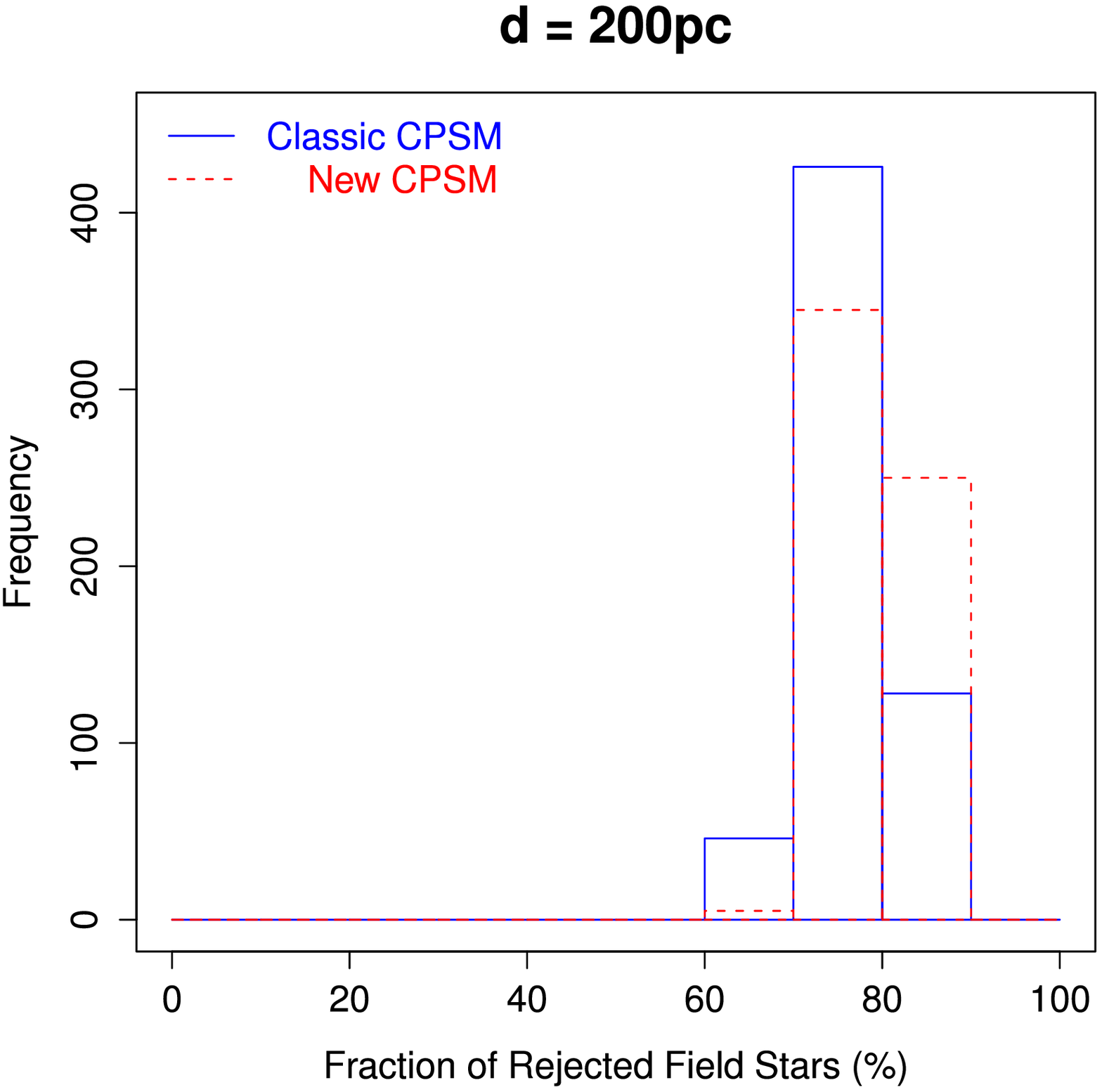}\vspace{0.4cm}
\includegraphics[width=8cm]{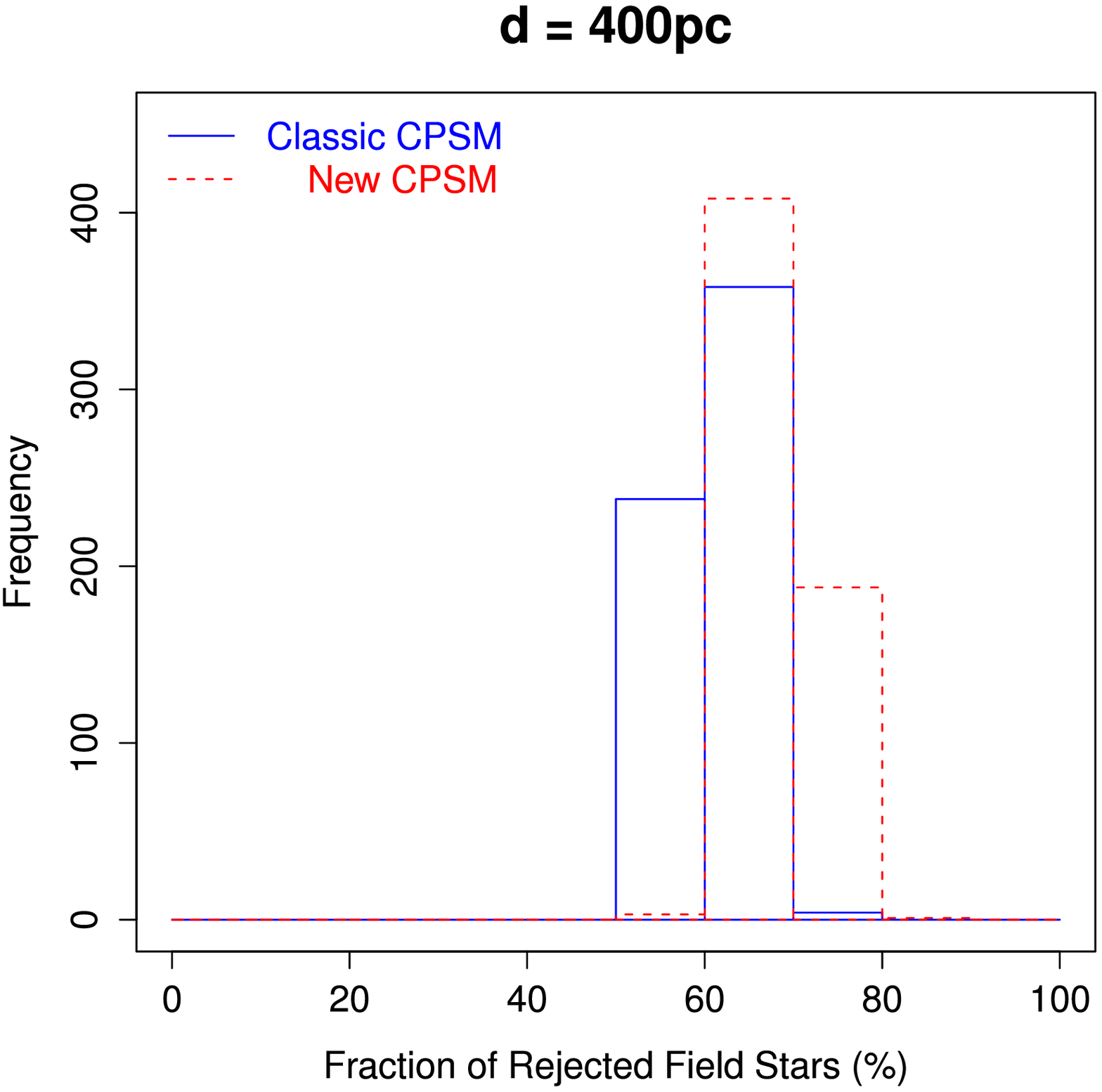}\vspace{0.4cm}
\caption{Fraction of rejected field stars for both CPSMs in the 1800 MC simulations. The field population is assumed to be at the same distance as the moving group.}
\end{center}
\label{fig6}
\end{figure}

\begin{figure}[!htp]
\begin{center}
\vspace{0.2cm}
\includegraphics[width=8cm]{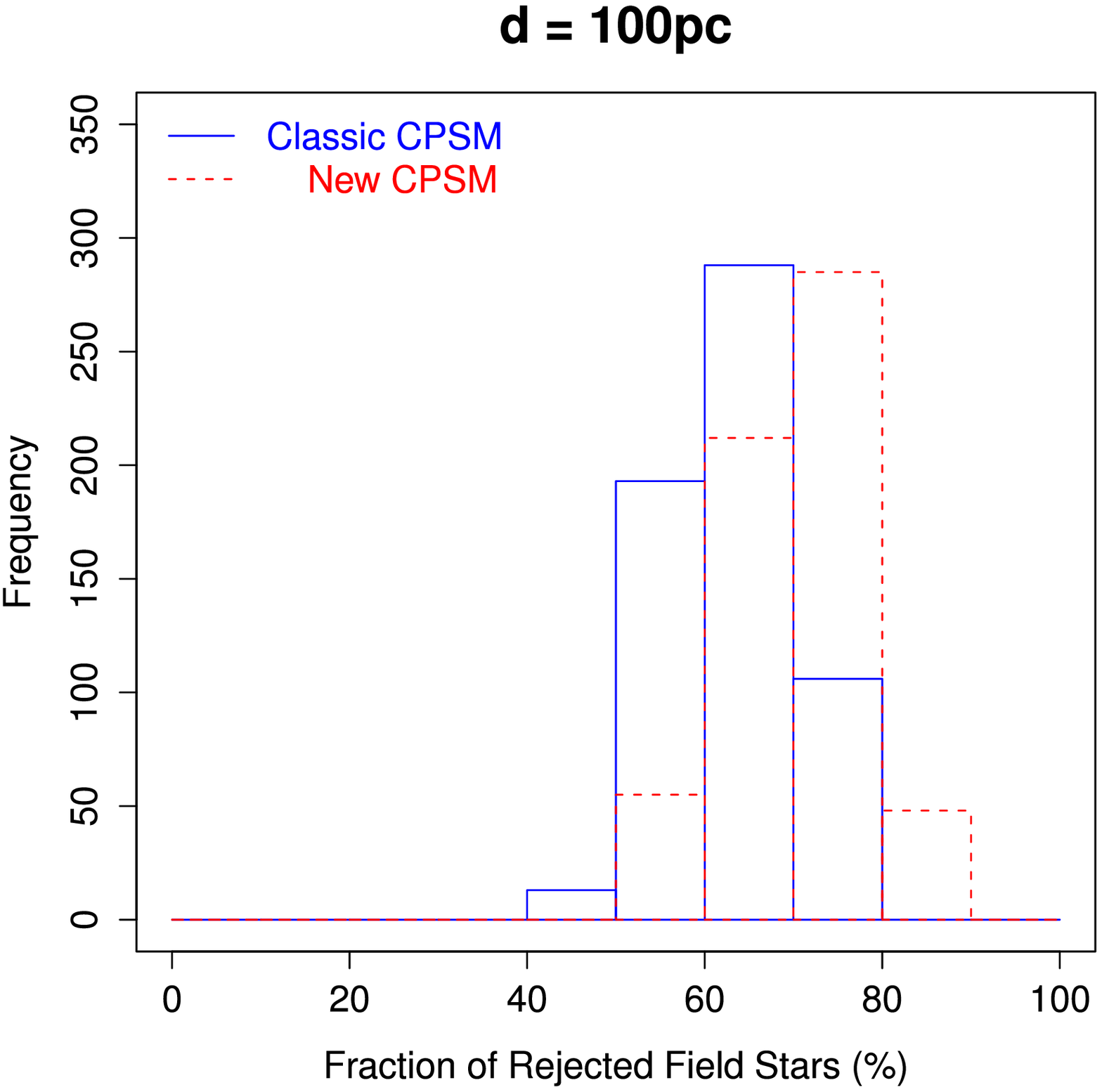}\vspace{0.4cm}
\includegraphics[width=8cm]{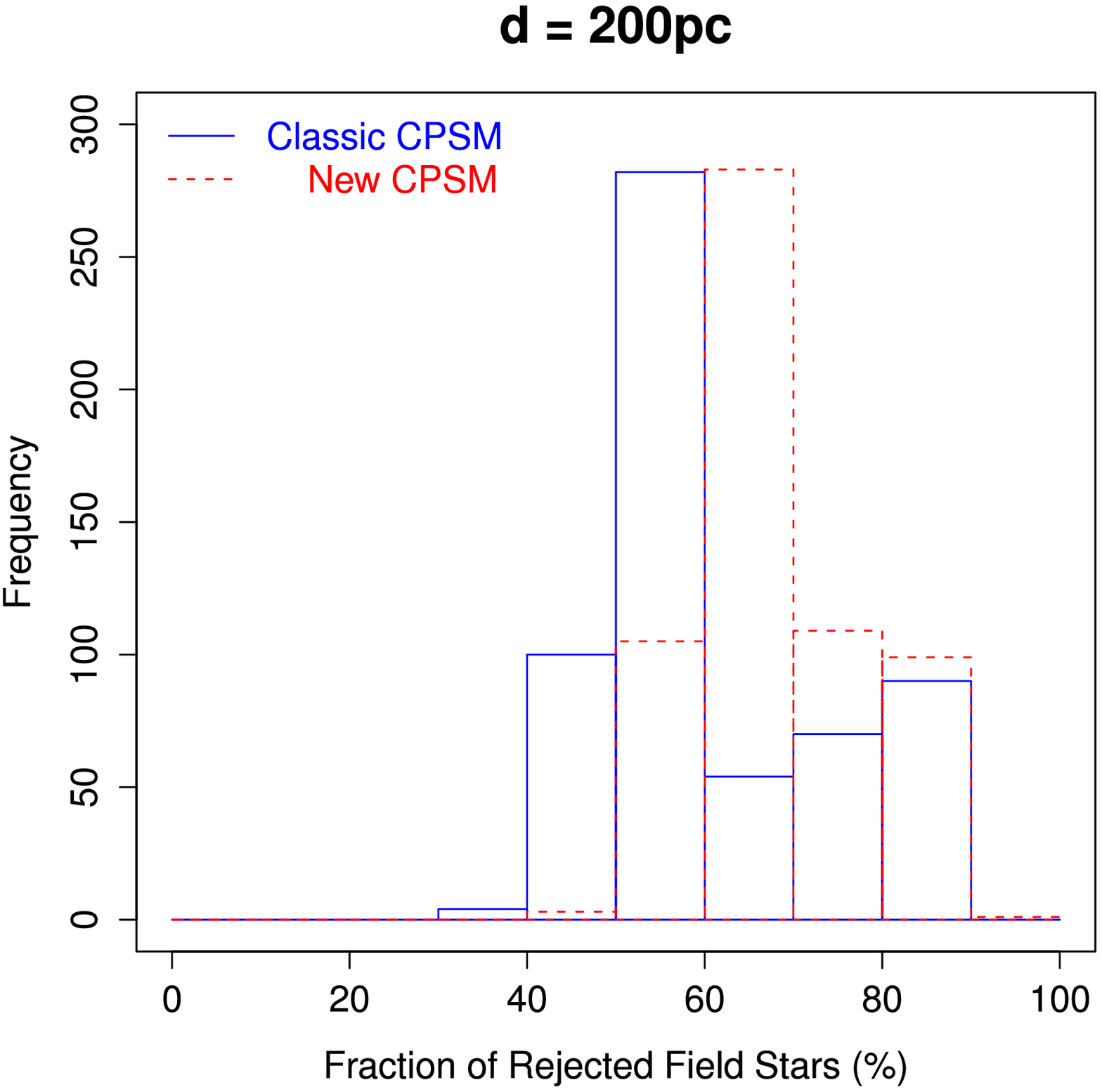}\vspace{0.4cm}
\includegraphics[width=8cm]{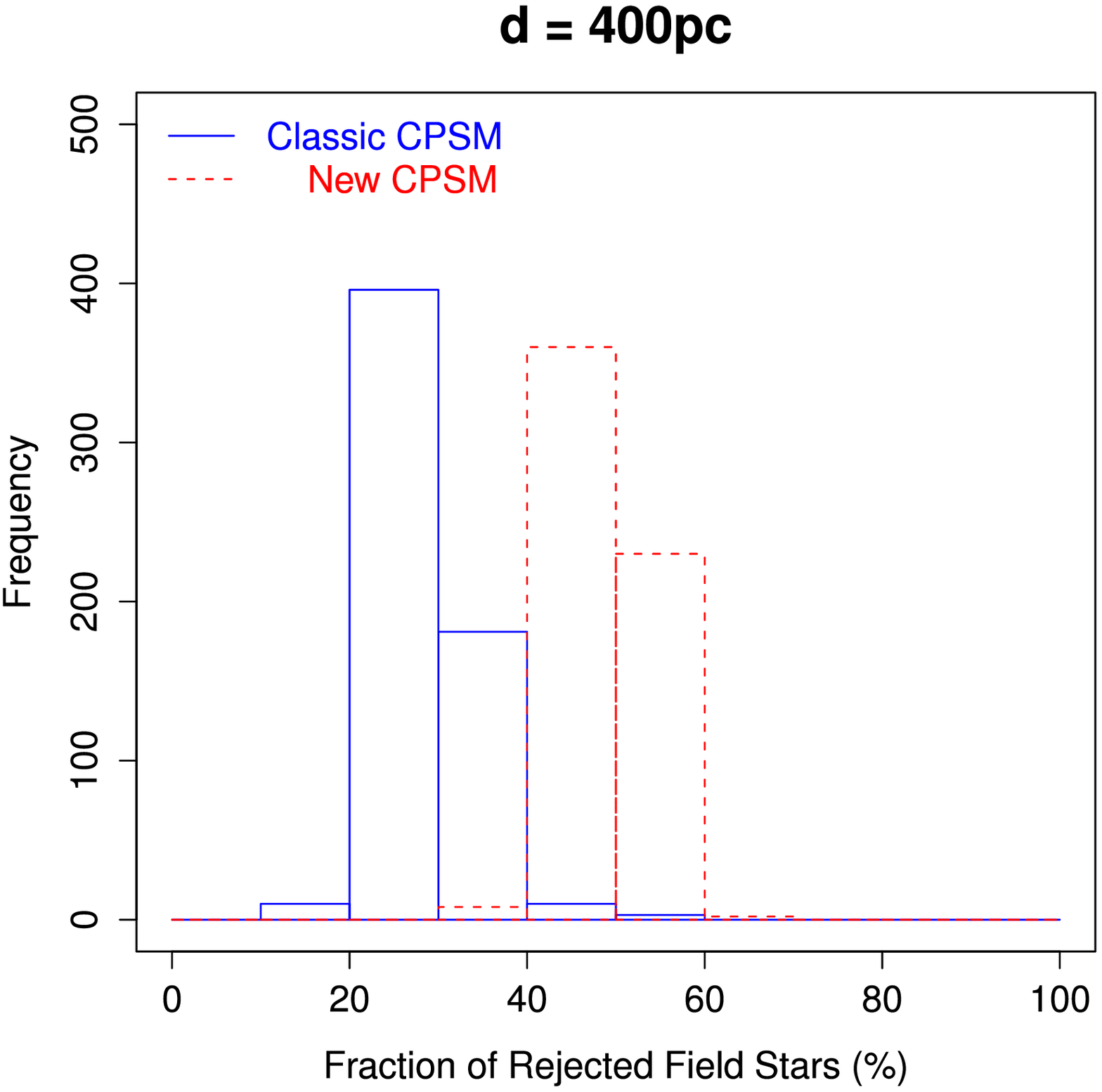}\vspace{0.4cm}
\caption{Fraction of rejected field stars for both CPSMs. The results summarize 1800 MC simulations. The field population consists of background stars at twice the distance of the moving group.}
\end{center}
\label{fig7}
\end{figure}

\subsection{Convergent point analysis}

As discussed in Sect.~4.2 the CP method is able to identify the moving group with a contamination of field stars that amounts to $\sim 20\%$ of the \textit{total} field population in the sample. Although the method retrieves a high fraction of cluster members, the CP is shifted from its true position due to the contribution of the remaining field stars in the sample. To better investigate the CP solution itself it is necessary to work on a sample where the moving group is the dominating population. In the following, we set $N_{MG}=80$ and $N_{field}=20$ in all configurations (A,B,C) of Table~1, construct our samples as described in Sect.~4.1, and present the results of CP analysis.

\subsubsection{The effect of velocity dispersion on CP positions}

As explained earlier, the velocity dispersion of the cluster prevents the proper motion vectors from pointing exactly towards the CP. Here we investigate how internal velocity dispersion affects both CPSMs by comparing the CP positional accuracy found for moving groups with different values of velocity dispersion. To do so, we construct three sets of synthetic data as described above, with $\sigma_{v}=$1.0, 2.0, and 3.0~km/s. We display in Fig.~5 the CP positions obtained with both CPSMs, overlaid on contour lines of constant $X^{2}$ values for a cluster with no internal velocity dispersion, centered on the same position and distance as the synthetic groups with nonzero internal dispersion velocities. The results shown in Fig.~5 are for a moving group with $(l,b)=(180^{\circ},0^{\circ})$ and $d=$100 pc. We find that the CP positions derived by both CPSMs are distributed along an elongated structure. The great circle on which the CP lies is defined by the proper motion vectors of group members, however the precise location of the CP over this great circle appears uncertain. As seen in Sect.~3, when strict convergence to the CP occurs, the individual poles of group members lie on a common plane that defines the polar great circle in the celestial sphere. Because of internal motions within the group, the poles are scattered around the polar great circle, and the precise position of the CP is not well defined, which explains the elongated structure observed in Fig.~5 \citep[see also][]{Bertiau1958}. When comparing both CPSMs, we find that they perform similarly and that the dispersion of CP positions always grows with the increasing velocity dispersion within the group.

\subsubsection{Precision of the CP position derived by each CPSM}

We now discuss the CP position uncertainties as derived by the covariance matrix. We compare the CP precision of both CPSMs in terms of the quantity $\Delta\sigma_{cp}=\sigma_{cp}^{classic}-\sigma_{cp}^{new}$, where $\sigma_{cp}$, the combined uncertainty of the CP position $(l_{cp},b_{cp})$ for each method, is given by
\begin{equation}\label{eq.26}
\sigma_{cp}=\sqrt{\sigma_{l_{cp}}^{2}+\sigma_{b_{cp}}^{2}}.
\end{equation}
The results of 1800 MC simulations for six values of the Galactic longitude (see above) and three different distances are shown in Fig.~6. When $\Delta\sigma_{cp}$ (normalized in order to represent only relative values between both CPSMs) is positive, the CP position uncertainties derived by the classic CPSM are higher than the ones obtained with the new CPSM. We find that the new CPSM returns a more precise CP solution for 95\% of the 1800 simulations. On the other hand, we also observe that the number of cases where $\Delta\sigma_{cp}< 0$, (i.e., where uncertainties in the classic CPSM are lower than in the new CPSM) grows with increasing distance of the moving group. In other words, the new CPSM is less precise than the classic CPSM at large distances, where the higher degree of concentration of the stars affects the determination of the CP position, defined as the intersection of the individual great circles of group members.

\begin{figure}[!htp]
\begin{center}
\vspace{0.2cm}
\includegraphics[width=7.7cm]{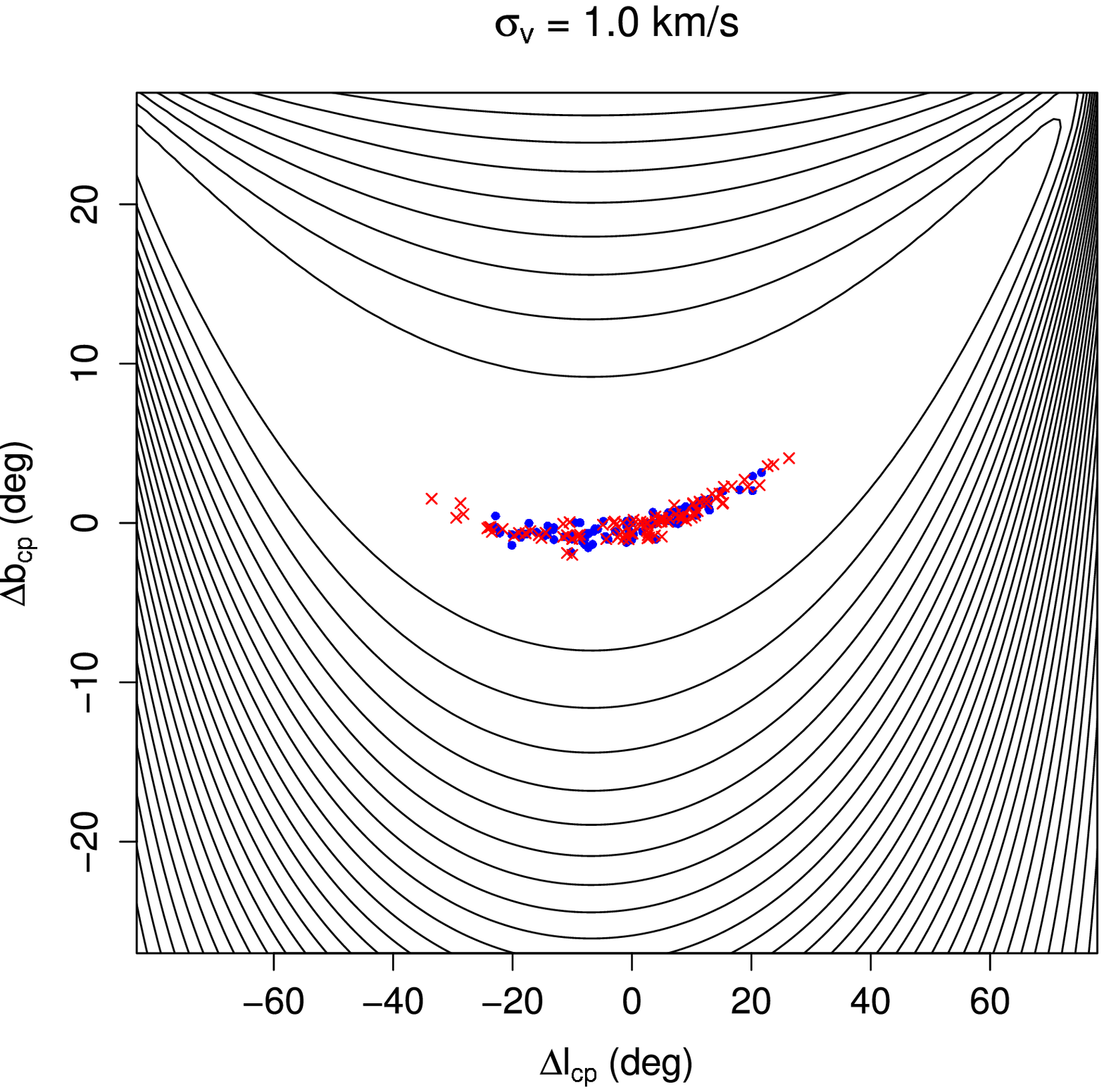}\vspace{0.4cm}
\includegraphics[width=7.7cm]{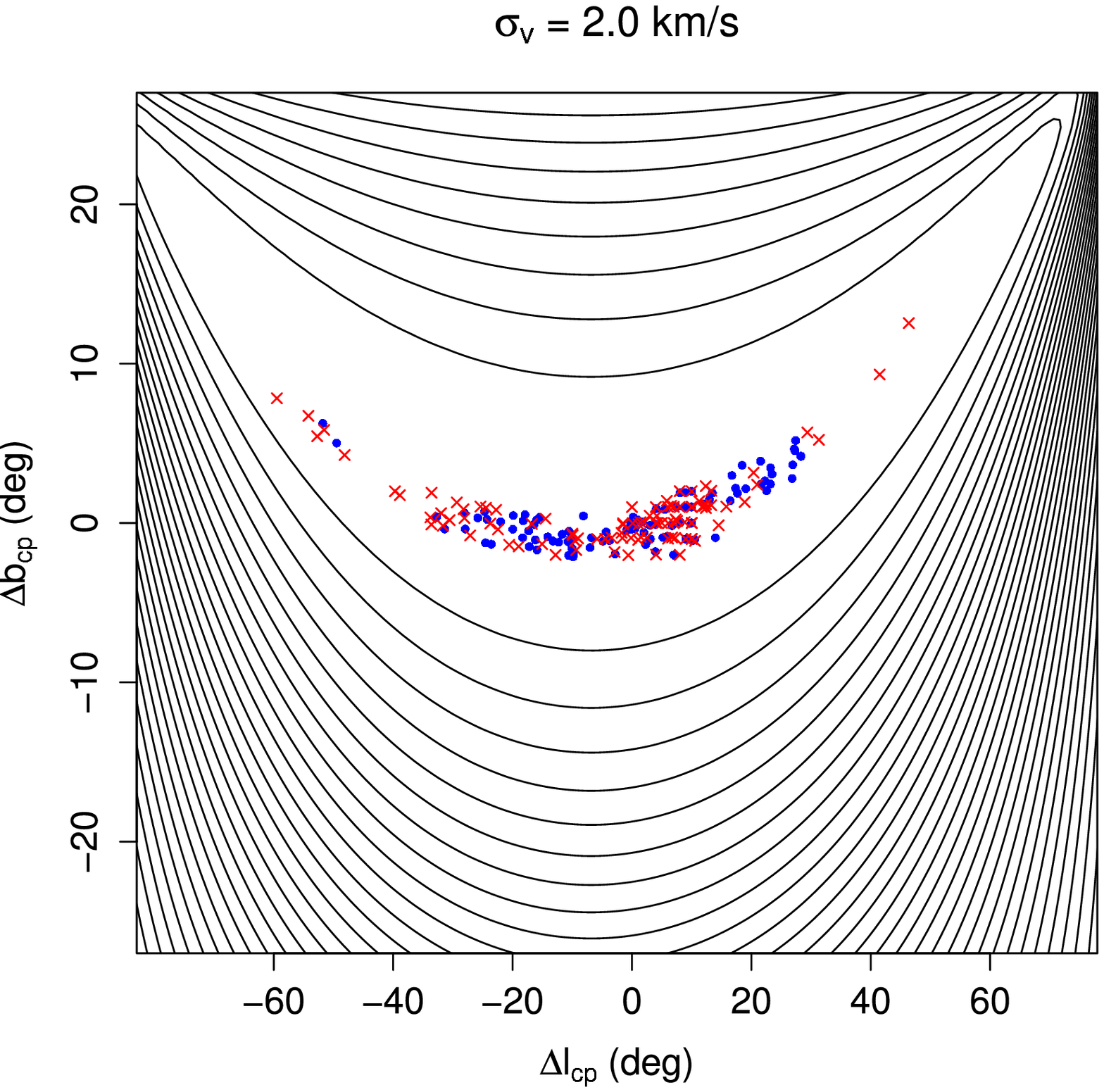}\vspace{0.4cm}
\includegraphics[width=7.7cm]{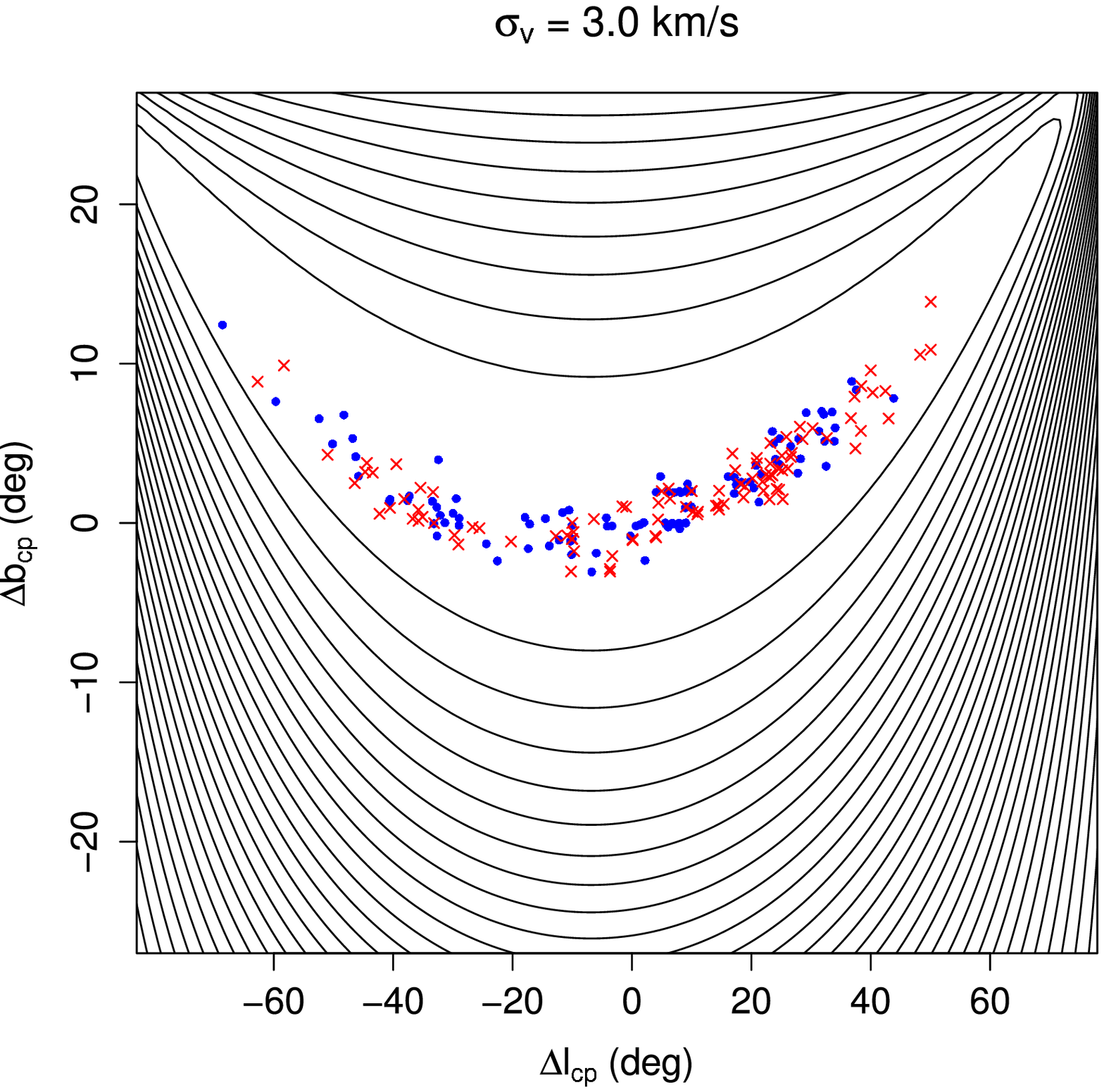}\vspace{0.4cm}
\caption{Relative position $(\Delta l_{cp},\Delta b_{cp})$ of the CP for 100 simulations and three different velocity dispersions. Blue dots and red crosses denote, respectively, the results of the classic and new CPSMs. The contours indicate $X^{2}$ levels linearly spaced by $\Delta X^{2}=2000$ for a cluster with the properties given in the text. The relative position of the computed CP $(l_{cp}^{sim},b_{cp}^{sim})$ for each simulation is given by $\Delta l_{cp}=l_{cp}^{sim}-l_{cp}$ and $\Delta b_{cp}=b_{cp}^{sim}-b_{cp}$.}
\end{center}
\label{fig3}
\end{figure}

\begin{figure}[!htp]
\begin{center}
\vspace{0.2cm}
\includegraphics[width=8cm, angle=0]{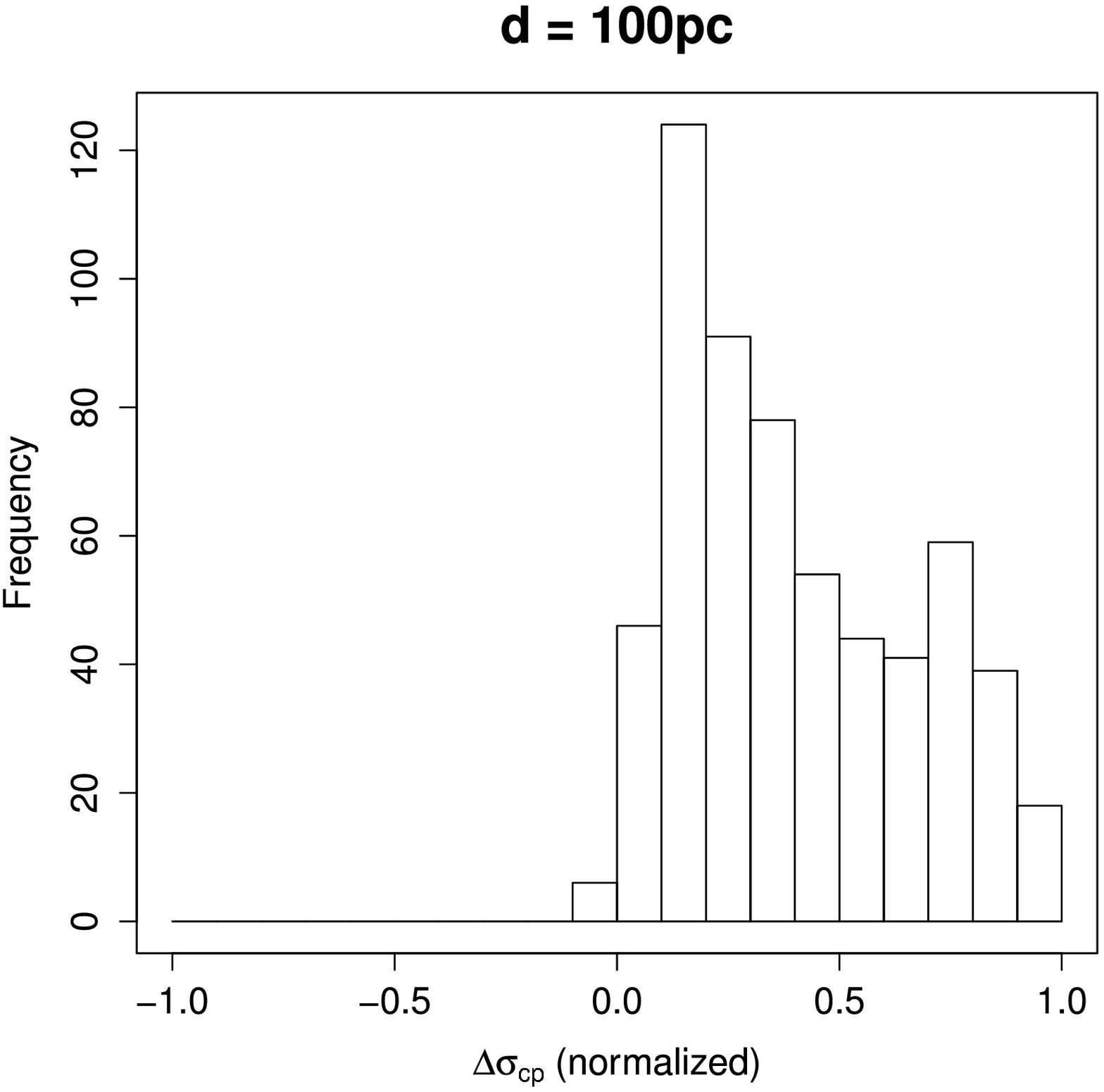}\vspace{0.4cm}
\includegraphics[width=8cm, angle=0]{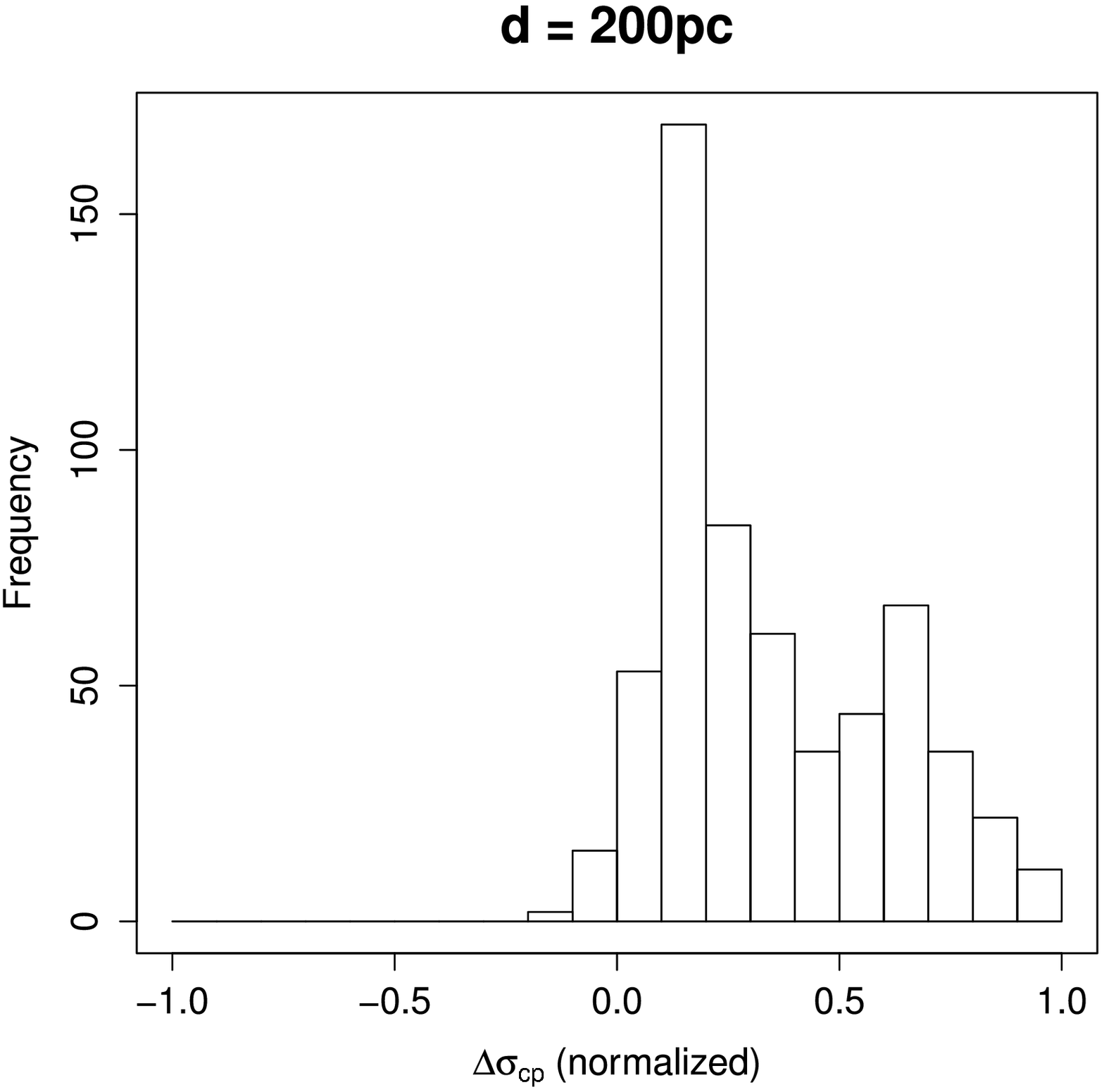}\vspace{0.4cm}
\includegraphics[width=8cm, angle=0]{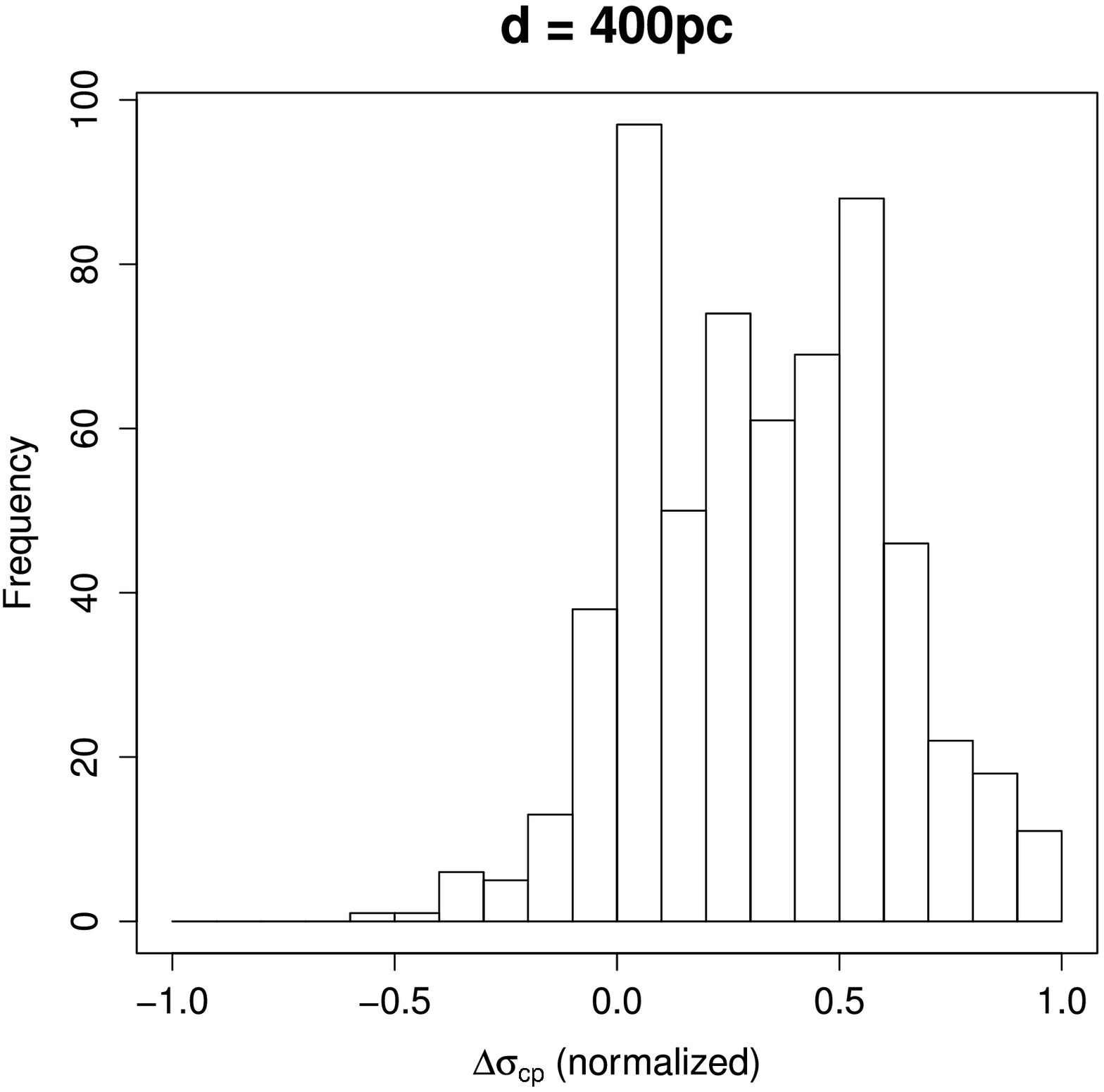}\vspace{0.4cm}
\caption{Comparison of the CP errors between both methods in the 1800 MC simulations of clusters located at three different distances.}
\end{center}
\label{fig4}
\end{figure}

\subsubsection{Role of moving group parameters on the solution accuracy of the new CPSM}

The precision of the CP position is also influenced by several other parameters such as observational errors on proper motions, number of moving group members, and angular distance from the moving group to the CP. The most efficient way to investigate the effect of each parameter is to consider an ideal cluster model in which the stellar velocity components  only result from the streaming motion of the cluster (i.e.,  the internal velocity dispersion is zero and the effects of Galactic rotation are not considered). We find, as expected, that for an ideal moving group both CPSMs return the same solution, with $X^{2}=0$ and $\epsilon = 1$. In the following, we focus on the role of each parameter for the results obtained with the new CPSM. We emphasize that the qualitative results of this investigation, obtained by using an ideal synthetic moving group, depend neither on the specific cluster position $(l,0^{\circ})$ nor on its distance. However, the amplitude of the CP uncertainties as given in Fig.~7 are specific to configuration \#A of Table~1 with the cluster centered at $(180^{\circ},0^{\circ})$, which we present here as an example to show the relative importance of each moving group parameter in the CP solution.

We first investigated the influence of proper motion errors on the CP position precision. We constructed synthetic samples of an ideal moving group (as defined above) by varying the relative error on proper motion data (in each component) from 5\% to 100\%. The results are presented in Fig.~7a and show that the errors on the CP position grow linearly with the increasing errors on proper motions. 

We now come back to the distance of the moving group and its effect on the CP position precision. As mentioned above, we expect the distance to influence the CP position errors since stars at greater distances have smaller proper motions, and the angular size of the group changes with respect to the distance. The question arises whether the precision of the CP is more affected by the deterioration of proper motion data at larger distances or by cluster concentration. Our numerical simulations make it possible to separate both effects. To do so, we simulate different moving groups varying one of these quantities while the other remains fixed. The results are shown in Figs.~7b and c. We find that cluster concentration is the dominant source of error on the CP position at large distances. 

We investigated the effect of the number of group members used for the CP position determination. Figure~7d shows the results of this computation, where we varied the number of moving group members from 10 to 300. We find that above $N=200$ no significant improvement in the CP position precision is achieved.

Finally, we seek to determine the effect of the angular distance from the moving group to the CP. To do so, we construct synthetic data sets by varying the Galactic longitude of the cluster center from $0^{\circ}$ to $360^{\circ}$ in steps of $0.15^{\circ}$. We consider the stellar three-dimensional velocity components to be isotropic, which fixes the CP position at $(45.00^{\circ},35.26^{\circ})$. Our results are shown in Figs.~7e and f. We find that moving groups displaced by $180^{\circ}$ have the same CP errors. This is because their relative position to the nearest solution (CP or DP) is the same. Those groups whose angular distance to the CP exceed $90^{\circ}$ are closer to the DP than to the CP. We recall that both solutions are always obtained with the same probability, so that the $X^{2}$ values are the same. We conclude that the angular distance between the moving group and the CP influences the precision of the solution. The least precise CP solution is obtained for moving groups displaced by $90^{\circ}$ with respect to the CP (or DP).

\begin{figure*}[!]
\begin{center}
\vspace{0.2cm}
\includegraphics[width=7cm]{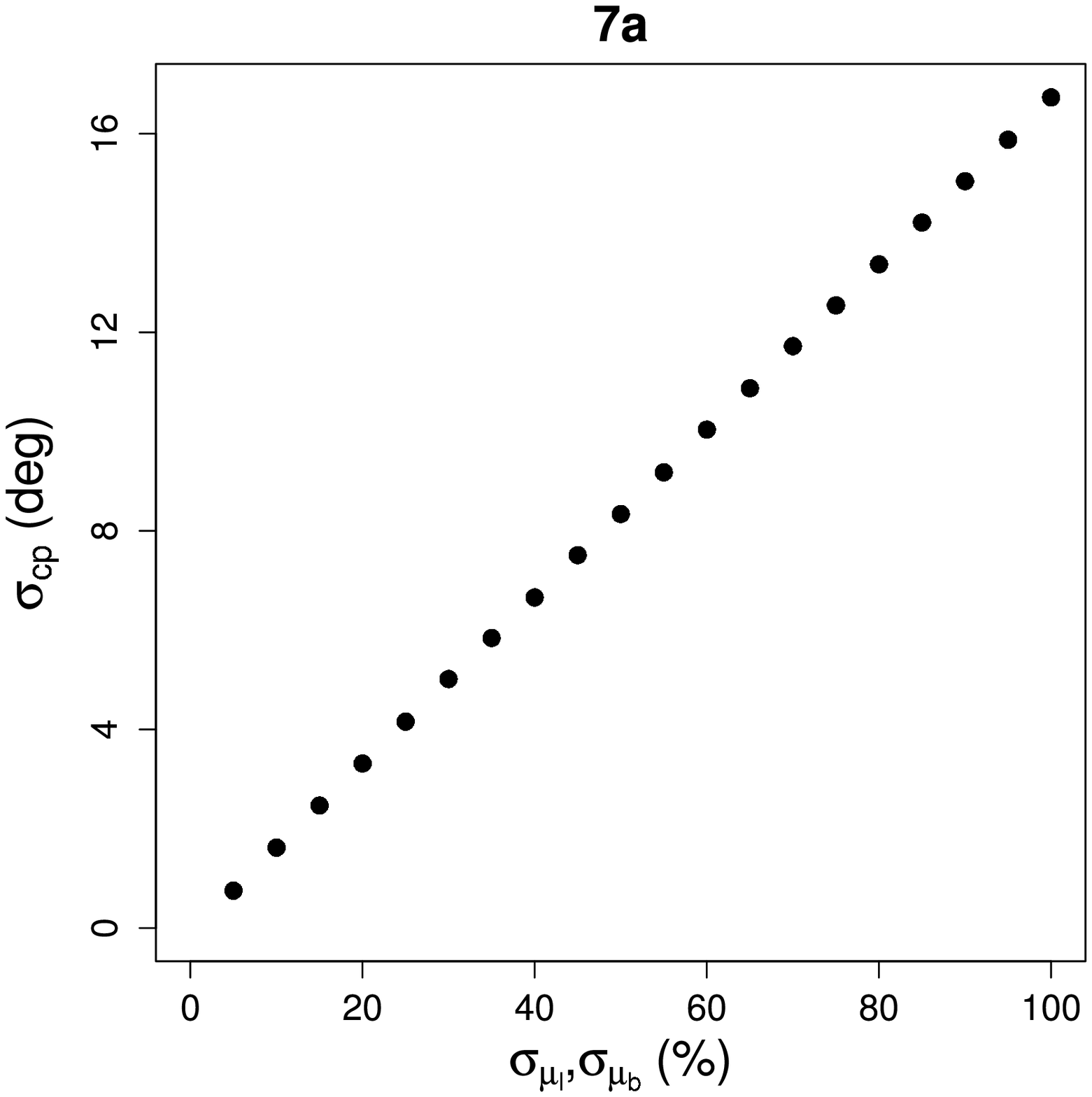}\vspace{0.4cm}\hspace{0.4cm}
\includegraphics[width=7cm]{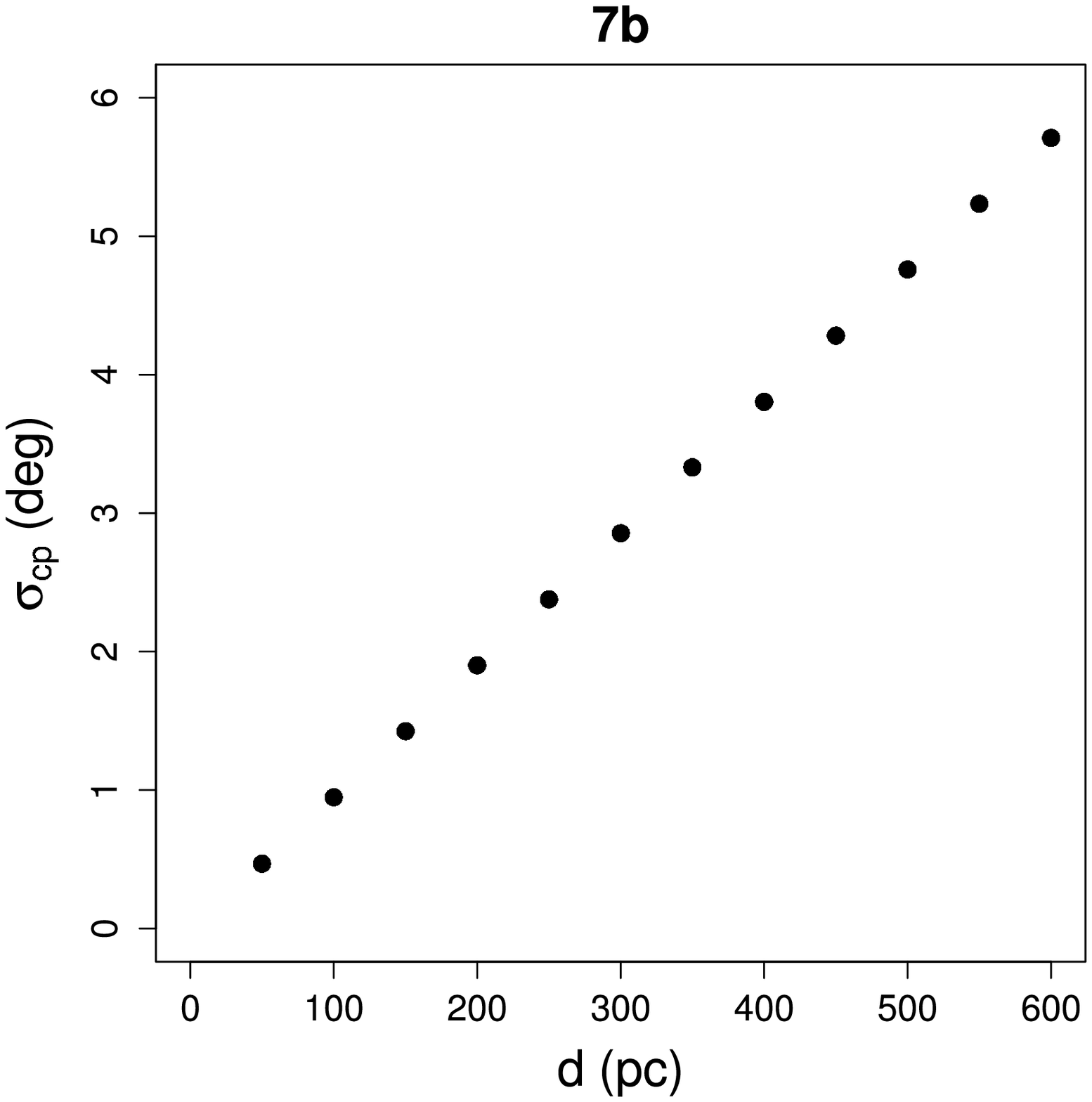}\vspace{0.4cm}\hspace{0.4cm}
\includegraphics[width=7cm]{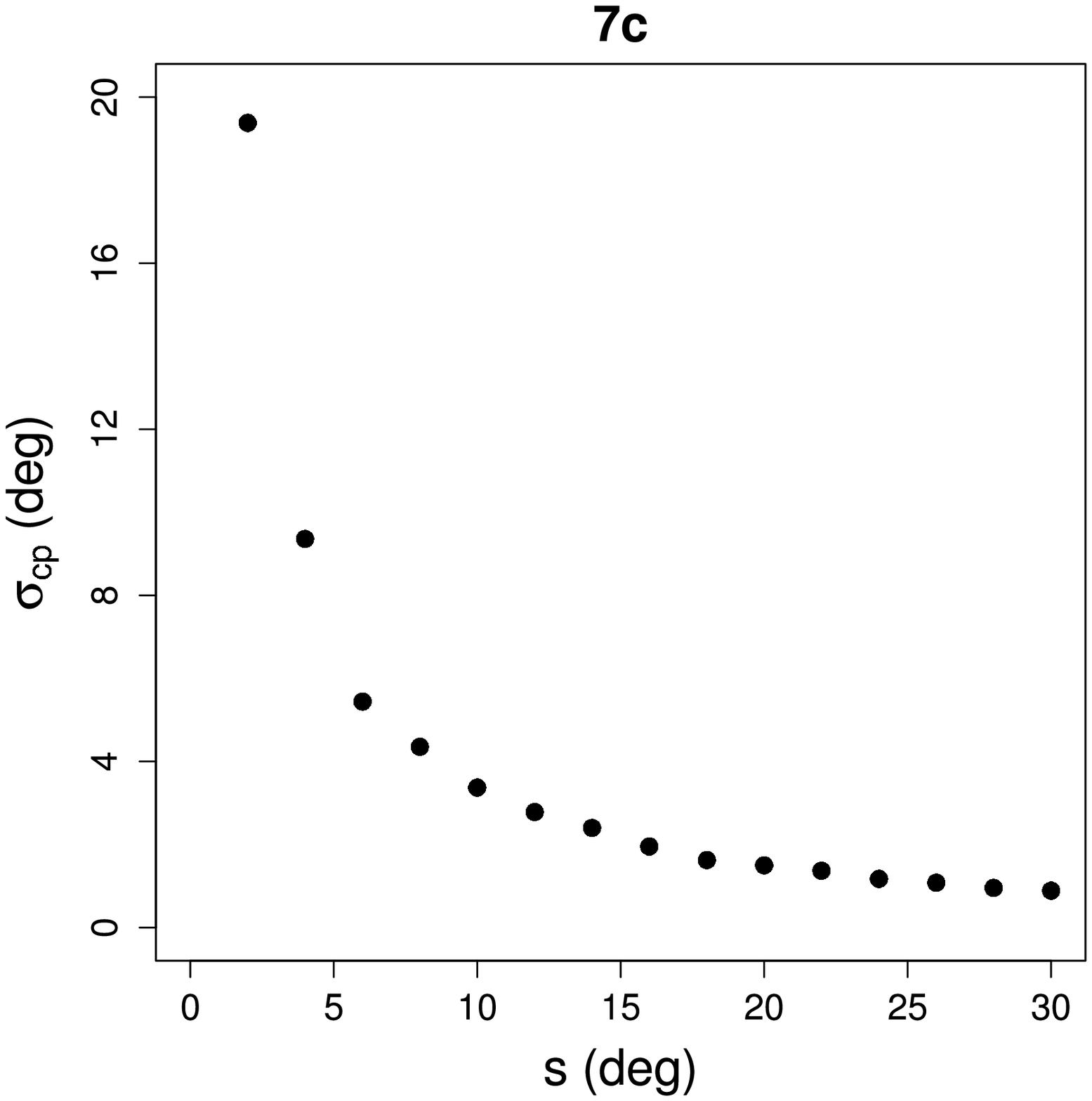}\vspace{0.4cm}\hspace{0.4cm}
\includegraphics[width=7cm]{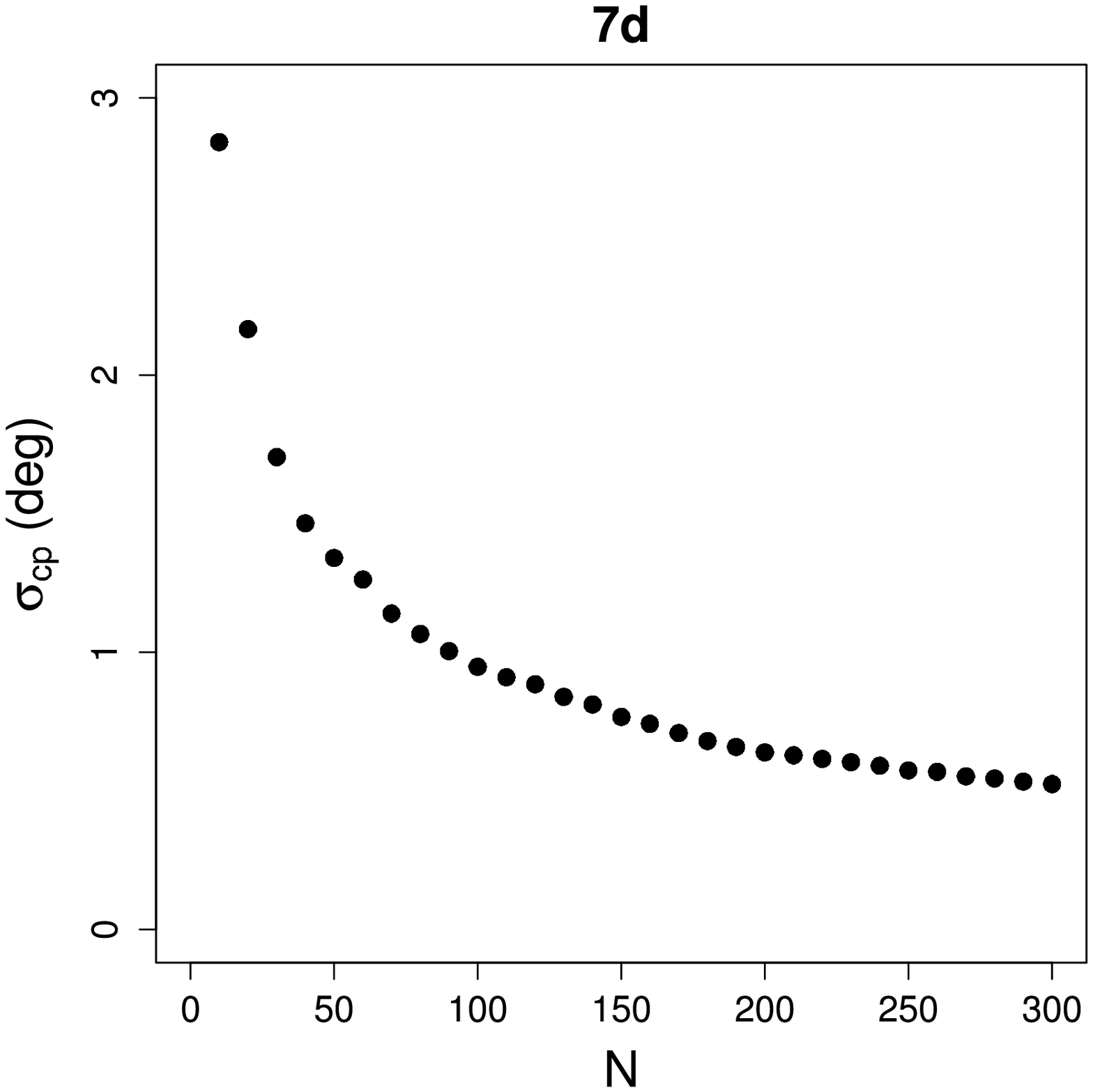}\vspace{0.4cm}\hspace{0.4cm}
\includegraphics[width=7cm]{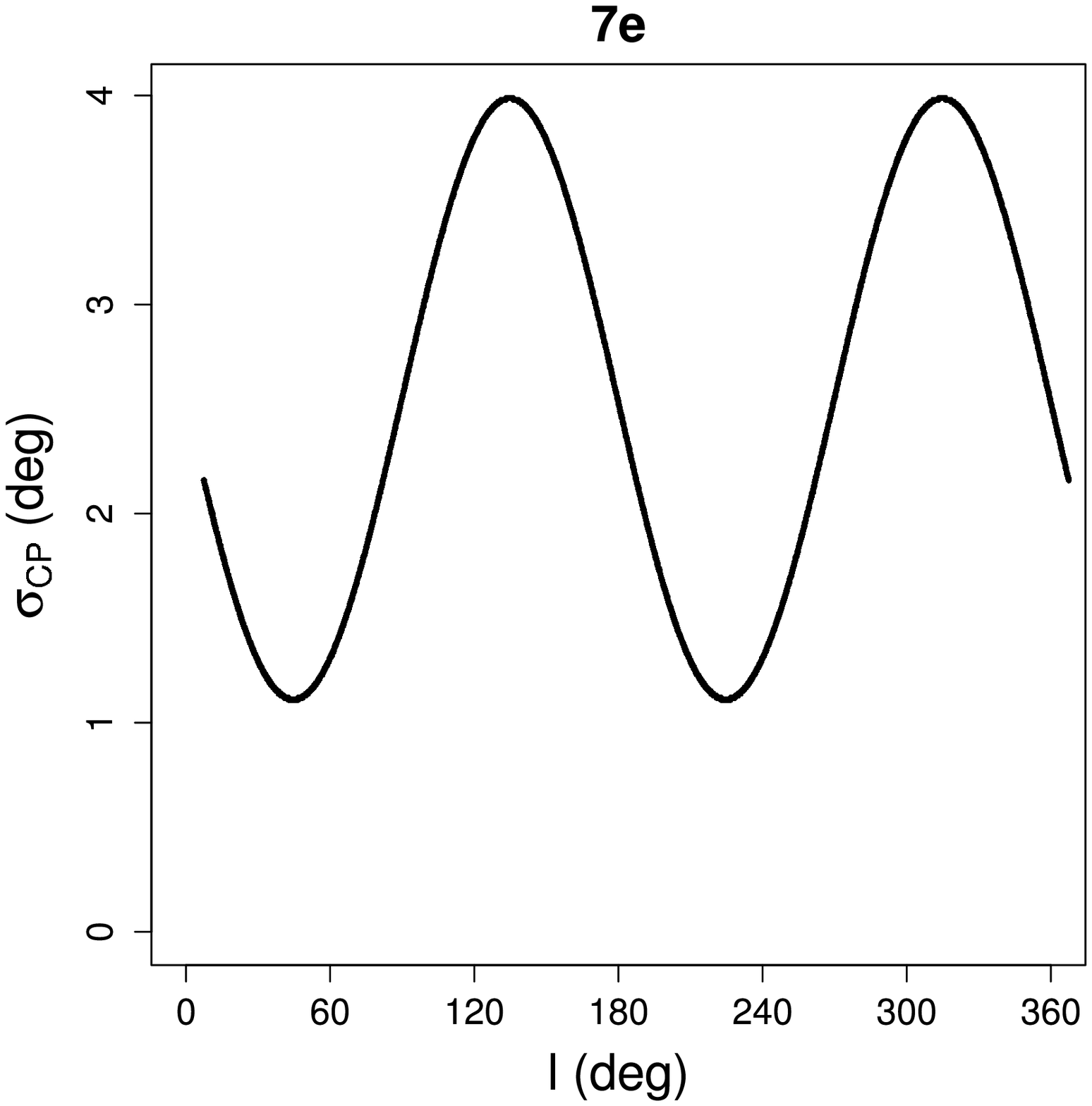}\vspace{0.4cm}\hspace{0.4cm}
\includegraphics[width=7cm]{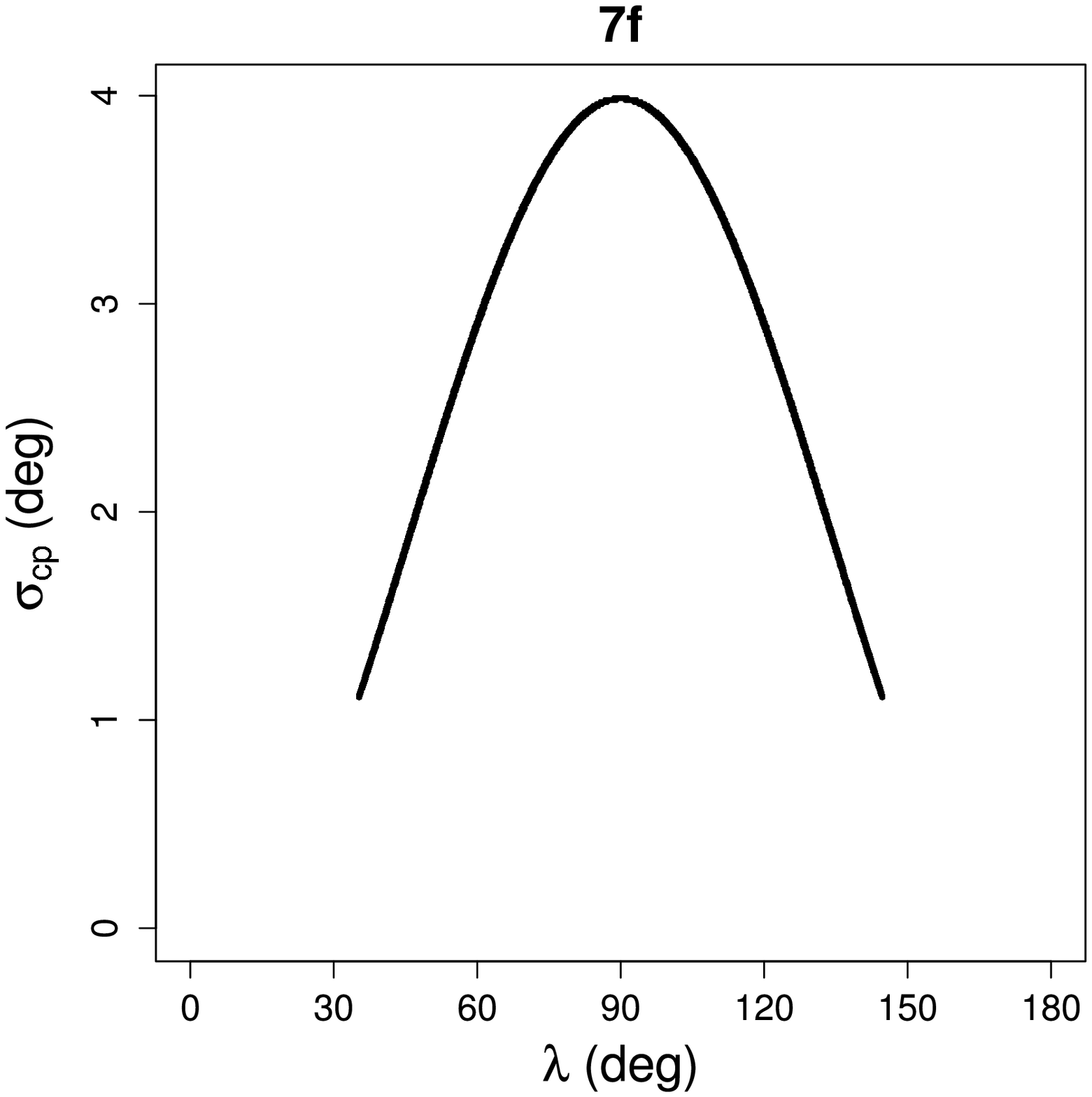}\vspace{0.4cm}\hspace{0.4cm}
\caption{Accuracy of the CP position as a function of (a) observational errors on proper motions, (b) distance with field size fixed, (c) field size of the group $(s\times s)$ with fixed distance, (d) number of  group members, (e) galactic longitude of the cluster with isotropic stellar velocity, and (f) angular distance $\lambda$ from the CP to the moving group.}
\end{center}
\label{fig5}
\end{figure*}

\subsection{Choice of $\epsilon_{min}$}

The final selection of moving group members depends mainly on the choice of $\epsilon_{min}$, the stop parameter that should allow finding the largest number of moving group members with the least contamination by field stars. So far, we have adopted in our new CP strategy the value of $\epsilon_{min}=0.954$, which was used by B99 in his implementation of the classic CPSM. This choice allows one to compare the performance of both CPSMs under the same conditions. To do so, we investigated whether the adopted value of $\epsilon_{min}$ by B99 is truly the optimum one for our new strategy. To do so, we ran the new CPSM on our synthetic data sets and varied the value of $\epsilon_{min}$ from 0.00 to 0.99. We then computed the fraction of true cluster members and field stars in the sample of stars selected by the new CPSM. The results are presented in Fig.~8  for a moving group with $(l,b)=(180^{\circ},0^{\circ})$ at $d=$100pc. Similar conclusions apply to all cluster configurations considered in this work. As already expected, the \textit{relative} fraction of group members in the sample of nonrejected stars is lower for the classic CPSM. This is because more field stars are accepted by this method. We verified that the number of rejected group members is equivalent for both CP methods. We find that for $\epsilon_{min}>0.954$ the number of moving group stars decreases while the fraction of field stars in the sample remains practically constant. Thus, we conclude that the value of $\epsilon_{min}=0.954$ can also be used in the new CPSM, since it offers the best compromise between the number of accepted group members and rejected field stars.

\subsection{Normality tests applied to the s distribution}

In Sect.~3.1we assumed that $s$ was normally distributed with zero mean and unit variance. Coming back to this point, we discuss the validity of this hypothesis. We carried out another set of Monte Carlo simulations with different configurations of cluster position, distance, angular size, and velocity dispersion in order to investigate whether the observed $s$ distribution is normal with zero mean and unit variance. The Kolmogorov-Smirnov normality test applied to the synthetic datasets reveals no significant differences between the simulations and the model. We present in Fig.~9 the results of this analysis for 1000 Monte Carlo simulations with different cluster configurations. We observe that the Kolmogorov-Smirnov test gives a p-value that is higher than the adopted significance level of 5\% for 99.6\% for our simulated moving groups. Thus, we conclude that the assumption of a normal distribution of $s$ is valid.

\section{Application to the Hyades}

Ever since the discovery that it is a nearby moving group, the Hyades open cluster has played a central role in astronomy in studies of Galactic structure, chemical evolution, distance calibration and stellar evolution models \citep[see for example][and references therein]{deBruijne(2001)}. Here we use the Hyades open cluster as a first application of our new CPSM to real data. We present the CP search using data from the \rm{\sc{Hipparcos}} catalog and compare our results with previous ones.

\subsection{Covariance matrix }

In general astrometric catalogs quote only standard errors on the observables. However, the \rm{\sc{Hipparcos}} catalog provides for each star a 5x5 covariance matrix for the measured astrometric parameters (position, proper motion, and parallax) and proper use of the full covariance matrix is mandatory. The new CPSM is adapted to consider the full covariance matrix of the \rm{\sc{Hipparcos}} data, and the specific formalism used for the propagation of errors is described in Appendix~C.

\begin{figure}[!htp]
\begin{center}\vspace{1cm}
\includegraphics[width=9cm]{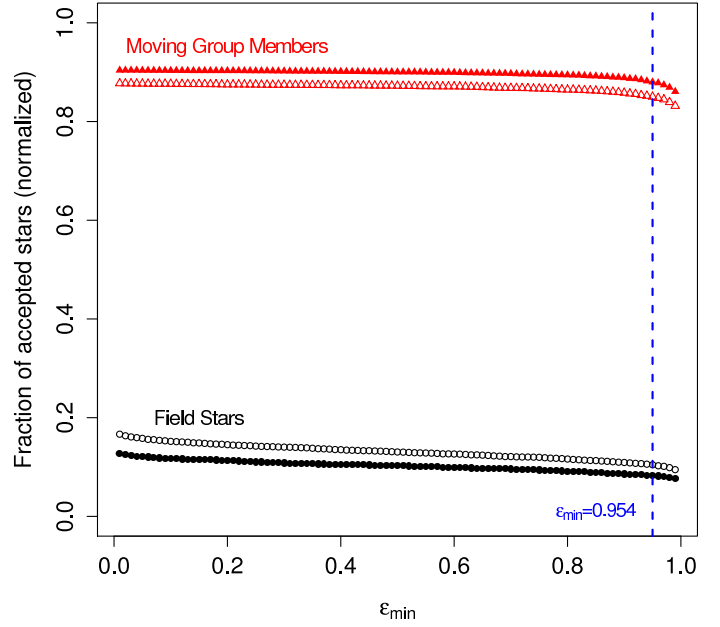}
\caption{Fraction of accepted group members and field stars in the sample as a function of the adopted value for $\epsilon_{min}$. Circles denote field stars and triangles cluster members. Open and filled symbols represent  the results obtained with the classic and new CPSMs, respectively. The vertical dashed line marks $\epsilon_{min}=0.954$. Each point represents the average value of 100 Monte Carlo simulations for a cluster with $(l,b)=(180^{\circ},0^{\circ})$ and $d=$100pc.\vspace{1.5cm}}
\end{center}
\label{fig8}
\end{figure}

\begin{figure}[!hbp]
\begin{center}
\includegraphics[width=9cm]{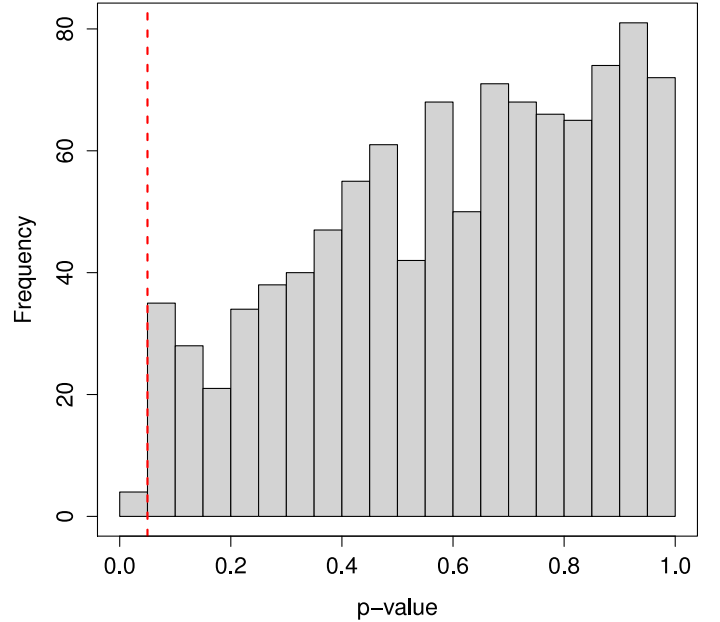}
\caption{Results of the Kolmogorov-Smirnov normality test applied to 1000 Monte Carlo simulations of moving groups with different configurations. The red dashed line marks the 5\% level of significance.}
\end{center}
\label{fig9}
\end{figure}

\vspace{0.5cm}
\subsection{CP analysis of the Hyades open cluster}

The sample of Hyades members that we use here consists of 218 stars from \rm{\sc{Hipparcos}} as given in \citet[][hereafter P98]{Perryman1998}, whose membership analysis was based on the three-dimensional velocity of the stars and the structure of the cluster. 

As already discussed by B99, a nonzero velocity dispersion is required in order to retrieve all cluster members. We adopt the value $\sigma_{v}=$2.0~km/s used by B99 and the mean distance of $d=$~46~pc as given in P98. After applying the new CPSM to the 218 members as the starting sample, the position of the CP is $(\alpha_{cp},\delta_{cp})=(97.69^{\circ},6.30^{\circ})\pm(0.89^{\circ},0.36^{\circ})$, and we retrieve 217 cluster members with a correlation coefficient of $\rho=-0.75$. Our solution is in good agreement with B99, $(\alpha_{cp},\delta_{cp})=(97.81^{\circ},6.74^{\circ})\pm(0.52^{\circ},0.21^{\circ})$ with 213 cluster members and $\rho=-0.84$. Another CP solution for the Hyades open cluster was found by \citet[][hereafter S91]{Schwan1991}, who obtained $(\alpha_{cp},\delta_{cp})=(97.68^{\circ},5.98^{\circ})\pm(0.42^{\circ},0.18^{\circ})$ using only 145 stars and proper motions from the FK5 and PPM catalogs. Figure~10 illustrates the proper motion vectors of the Hyades members and the derived CP coordinates. We find that only one star, HIP 28774, was rejected in our analysis as a nonmember of the Hyades open cluster. Its proper motion is given as $(\mu_{\alpha}\cos\delta,\mu_{\delta})=(-2.46,-7.00)\pm(18.43,9.99)$~mas/yr in the original \rm{\sc{Hipparcos}} catalog \citep[][hereafter HIP97]{HIP97} and as $(\mu_{\alpha}\cos\delta,\mu_{\delta})=(42.95,-2.16)\pm(14.68,8.04)$~mas/yr in the new \rm{\sc{Hipparcos}} reduction \citep[][hereafter HIP07]{HIP07}. These two values are different with low quality, as shown by the large uncertainties. The mean uncertainty on the proper motions of other cluster members is $\leq 2$~mas/yr in each component. Furthermore, the parallax changes from$\pi_{HIP97}=(12.81\pm12.80)$~mas to $\pi_{HIP07}=(39.18\pm11.54)$~mas, which reflects the uncertainties of the \rm{\sc{Hipparcos}} data for this star. That the new CPSM considered that HIP 28774 is not a Hyades member thus appears justified.

\begin{figure}[!htp]
\vspace{0.5cm}
\includegraphics[width=8.8cm]{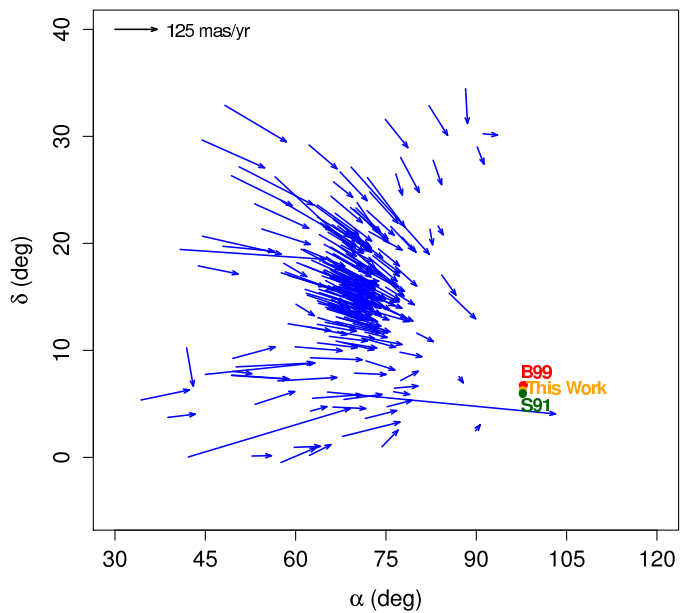}
\caption{Proper motion vectors of the Hyades members and the CP coordinates derived in different works. The CP error bars are too small to be seen.\vspace{1cm}}
\label{fig7}
\end{figure}

\subsection{Determination of individual parallaxes}

Once the CP coordinates of the Hyades have been determined, a final check and further extension of our CPSM lies in computing individual parallaxes and comparing it with HIP97 and HIP07. The determination of individual parallaxes in this work is restricted to group members whose radial velocity is known and given in Table~2 of P98. The individual parallaxes of group members are given by
\begin{equation}\label{eq.27}
\pi=\frac{A\,\mu_{\parallel}}{V_{r}\tan\lambda},
\end{equation}
where the corresponding quantities and units are defined in Sect.~2. This equation is also valid for linear expansion of the group mentioned before \citep[see][]{Blaauw1964}. The parallax uncertainty is derived by error propagation of this equation and takes the error budget of proper motions and radial velocities into account. 

We present in Fig.~11 a comparison of the parallaxes derived in this work with the ones given in HIP97 and HIP07. We find a slight bias when comparing our results with HIP97, which disappears when we compare our parallaxes with the presumably more accurate ones given in HIP07. Another dataset that we used to compare the individual parallaxes of this work are the secular parallaxes of Hyades cluster members given by \citet{deBruijne(2001)}. These secular parallaxes were calculated using proper motion data from the \rm{\sc{Hipparcos}} and \rm{\sc{Tycho2}} catalogs. The results of this comparison for each data set are shown in Fig.~12. We conclude that our results agree well with the more precise secular parallaxes of the Hyades open cluster.

\begin{figure*}[!htp]
\begin{center}
\vspace{1cm}
\includegraphics[width=0.49\textwidth]{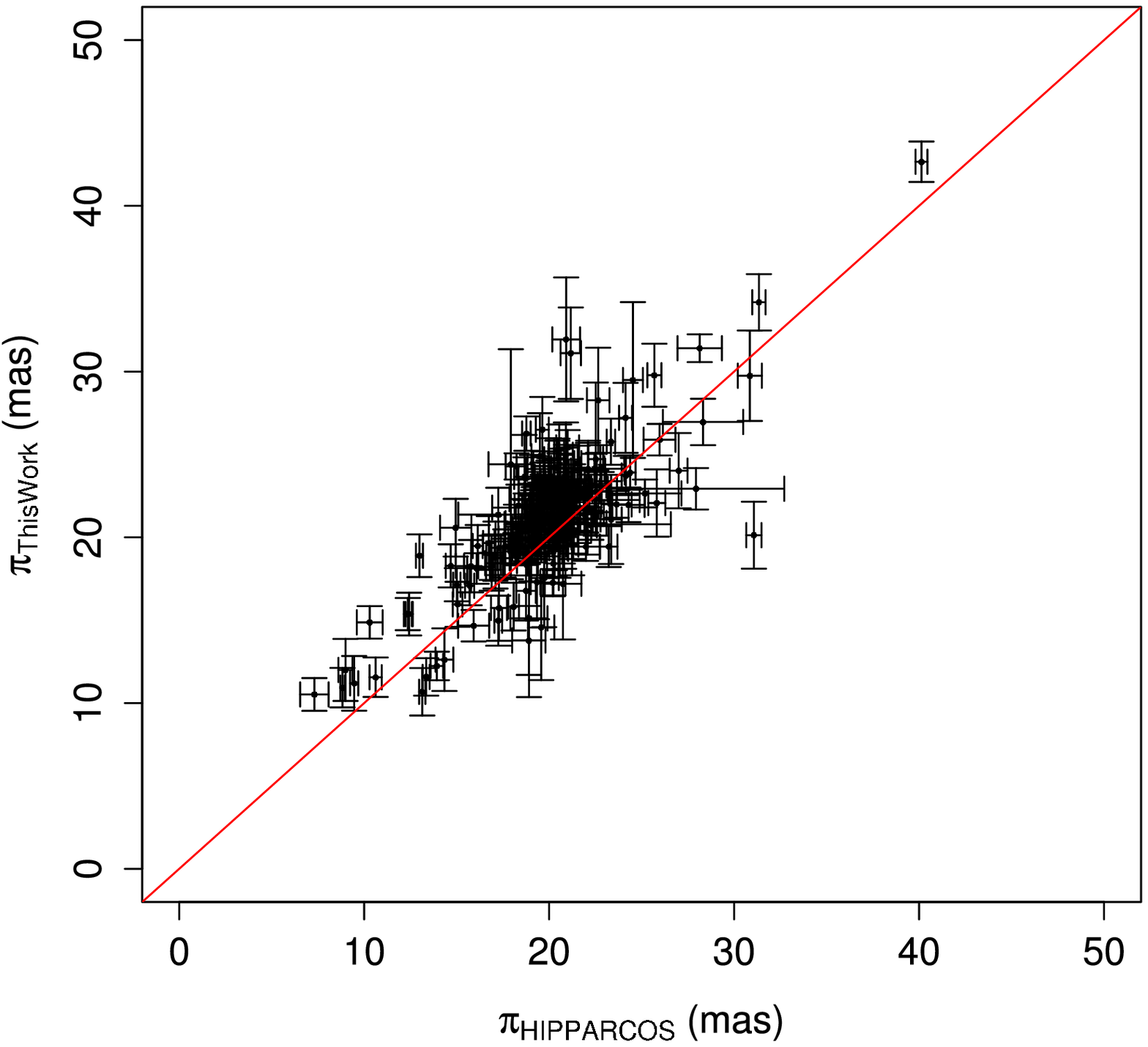}\hspace{0.2cm}
\includegraphics[width=0.49\textwidth]{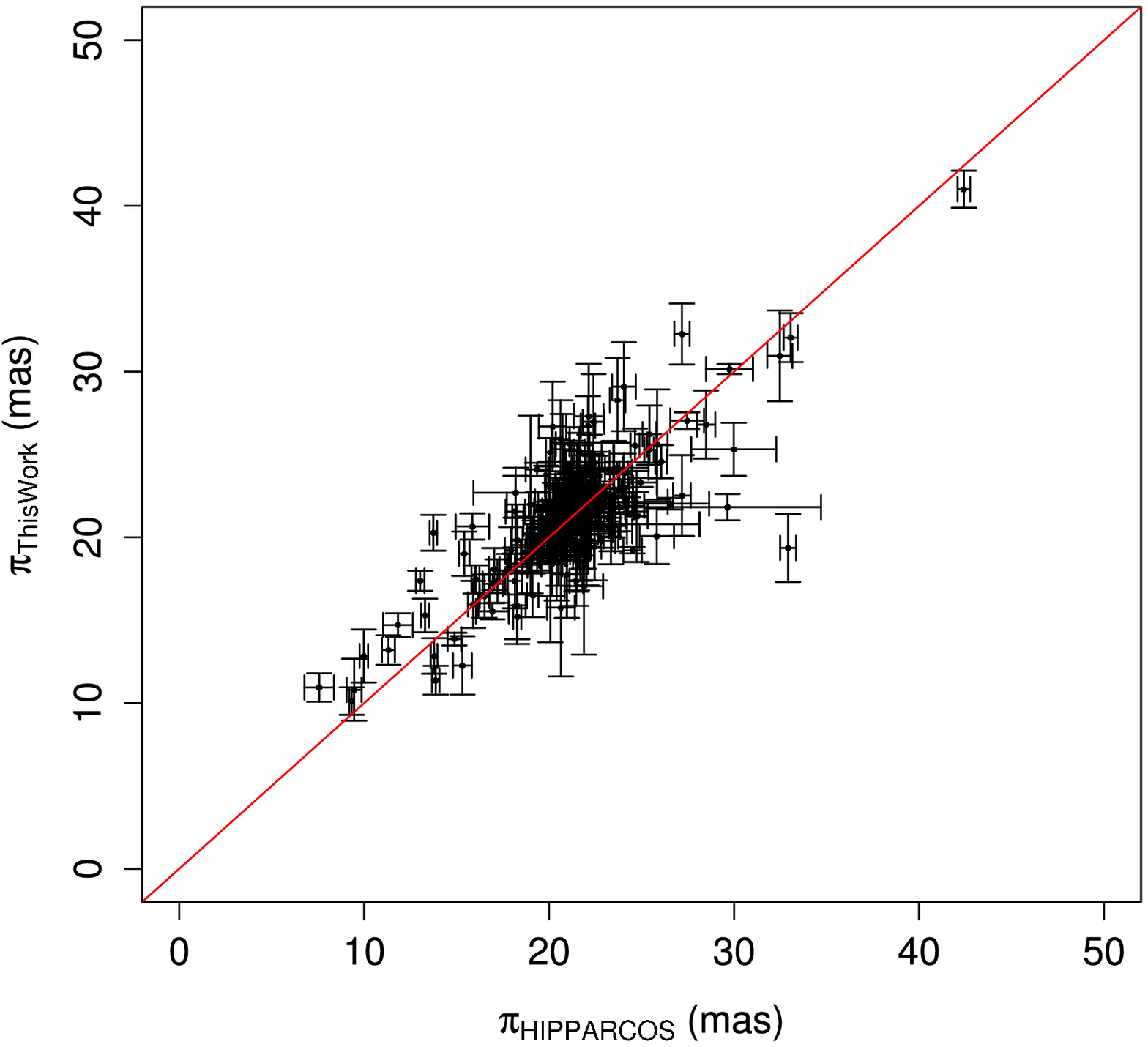}
\caption{Individual parallaxes of Hyades members derived in this work compared with their counterparts in HIP97 \textit{(left panel)} and HIP07 \textit{(right panel)}. The red line represents the expected distribution for equal results. The rms with respect to HIP97 is 3.62~mas and 3.34~mas for HIP07. The mean difference between the parallaxes derived in this work, and the ones in HIP97 and HIP07 are 1.18~mas and 0.02~mas, respectively. \vspace{1.0cm}}
\end{center}
\label{fig8}
\end{figure*}

\begin{figure*}[!htp]
\begin{center}
\vspace{1cm}
\includegraphics[width=0.49\textwidth]{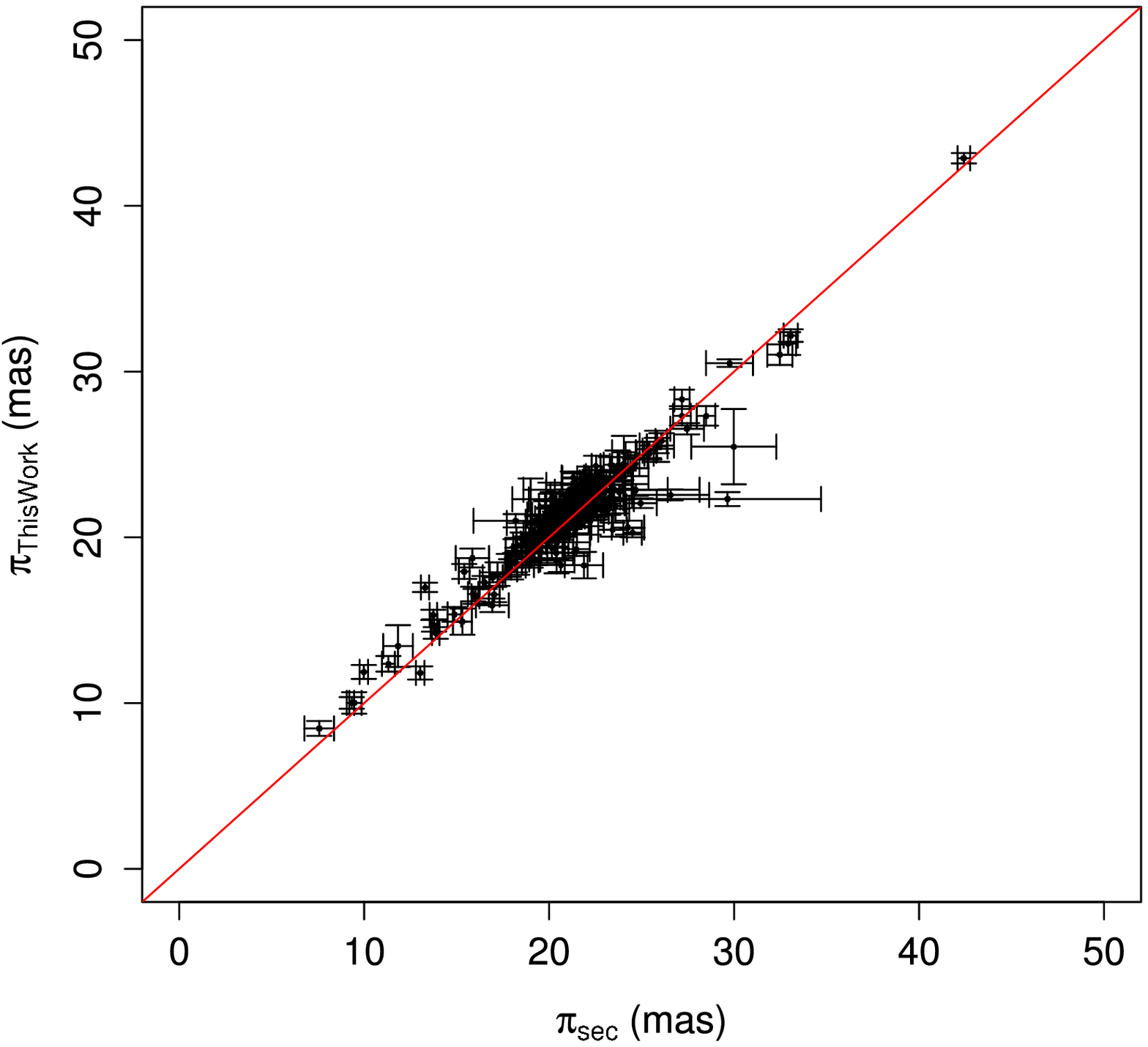}\hspace{0.2cm}
\includegraphics[width=0.49\textwidth]{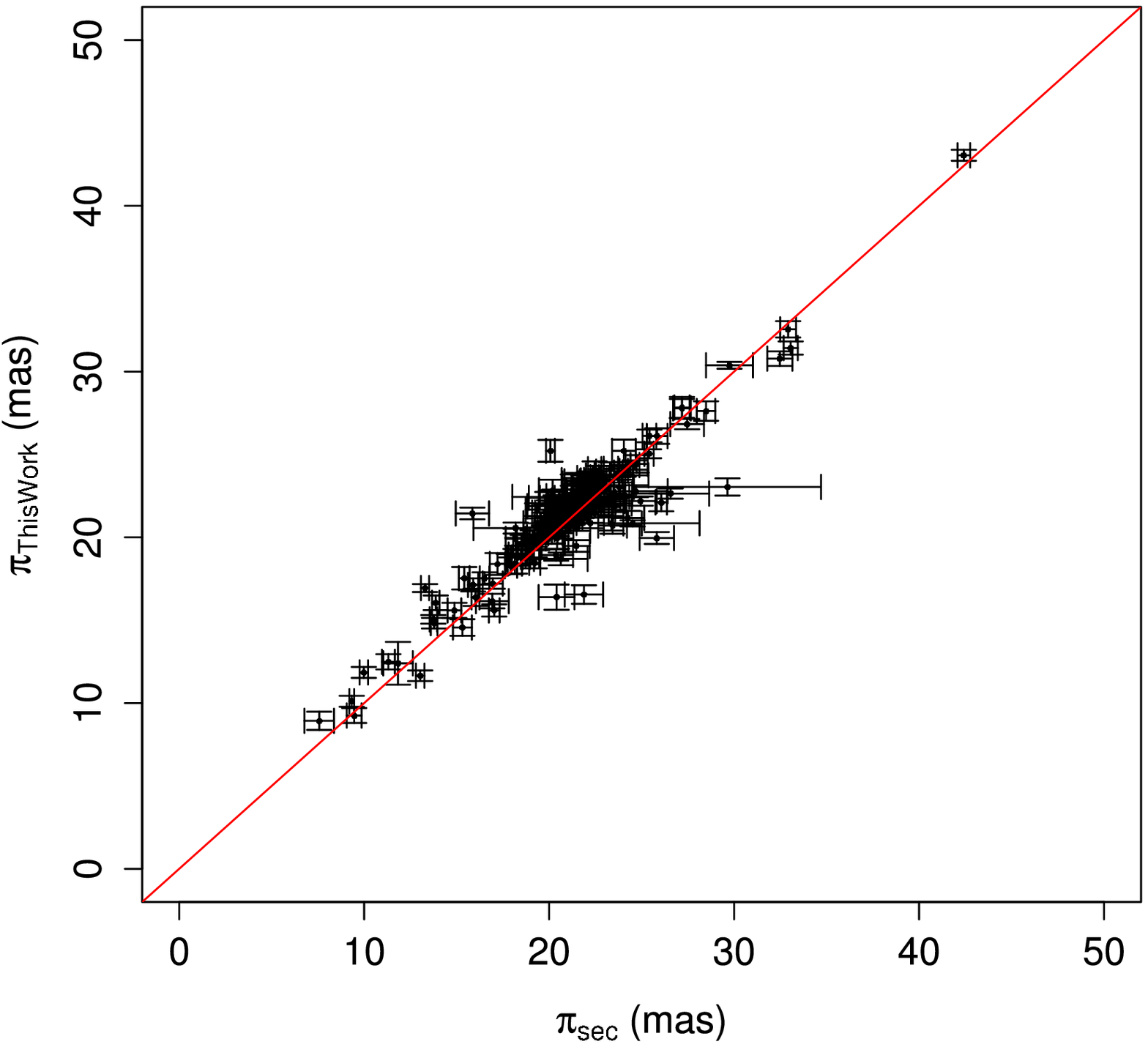}
\caption{Individual parallaxes of Hyades members derived in this work compared to the secular parallaxes of \citet[]{deBruijne(2001)} derived using proper motion data from the \rm{\sc{Hipparcos}} \textit{(left panel)} and \rm{\sc{Tycho2}} \textit{(right panel)} catalogs. The red line represents the expected distribution for equal results. The rms with respect to the \rm{\sc{Hipparcos}} secular parallaxes is 1.55~mas and 1.68~mas for \rm{\sc{Tycho2}} secular parallaxes, and the mean difference between the parallaxes is 0.23~mas and -0.27~mas.\vspace{1.0cm} }
\end{center}
\label{fig8}
\end{figure*}

\section{Summary}

We presented a new CPSM based on the idea of representing the stellar proper motions by great circles on the celestial sphere. The new CP method combines  (i) maximum-likelihood analysis to simultaneously determine the CP and select moving group members with (ii) a direct minimization routine used to return a more refined CP position and its uncertainties. Our new method allows us to correct for internal motions within the group by applying an individual correction for each star that depends on its proper motion. This procedure takes into account that the stars in the group are not located at the same distance.  

We performed extensive Monte Carlo simulations to test and compare our new CPSM with the classic CPSM regarding the selection of moving members and the accuracy of the CP solution. We investigated the effects of (a) the velocity dispersion of the cluster, (b) observational errors on proper motion, (c) cluster distance, (d) number of group members, and (e) angular distance to the CP on the CP solution. Our new CPSM returned a more precise CP solution than the classic CPSM for 95\% of the simulations. We verified that in the absence of velocity dispersion (ideal moving group) both methods exhibit the same performance. We also found that the new CPSM finds and eliminates more field stars than the classic CPSM. This situation is even more evident in the presence of background stars that generally have smaller proper motions. The new CPSM is able to retrieve more than 80\% of all cluster members with a contamination around 20\% of the total number of field stars at distances of 100~pc and 200~pc. At larger distances the efficiency of rejecting field stars decreases, but it is still higher than the efficiency of the classic CPSM. The definition of the $X^{2}$ function in the new CP method considers both the amplitude and direction of the stellar proper motion vector instead of only one directional component that removes the bias towards distant stars that is observed in the classic CPSM. Additional information (e.g. parallaxes and radial velocities) is required to eliminate the remaining field stars in the sample that were not rejected by the CPSM. 

The new CPSM is shown to work well when applied to the Hyades open cluster, and its results are agree well with previous determinations of the CP position. The individual parallaxes derived using our CP solution are fully consistent with the trigonometric parallaxes given in the \rm{\sc{Hipparcos}} catalog. Although our results agree well with data from both reductions, we find that our external precision is slightly better for HIP07. Our results are also in good agreement with the secular parallaxes found in the literature, which are more precise than the trigonometric parallaxes given in the \rm{\sc{Hipparcos}} catalog.

The original implementation of the new CPSM was adapted to take the full covariance matrix of the \rm{\sc{Hipparcos}} catalog into account. The same procedure can also be applied to the future data of the \textit{Gaia} mission, which will also be published with the full covariance matrix. The new CPSM will be used in forthcoming papers to investigate the kinematic properties of several young associations.

\begin{acknowledgements}
It is pleasure to thank Alberto Krone Martins and Michel Rapaport for helpful and stimulating discussions. We are also grateful to the referee for constructive comments that helped us to improve the manuscript. This research was funded by a doctoral fellowship from FAPESP and made use of the SIMBAD database operated at the CDS, Strasbourg, France.
\end{acknowledgements}

\newpage
\bibliographystyle{aa}
\bibliography{references.bib}

\clearpage
\onecolumn
\appendix

\section{Pole coordinates $(\alpha^{p},\delta^{p})$}

The rectangular coordinates of the normalized polar vector $\mathbf{p}$ are given as
\begin{equation}\label{eq.A1}
\mathbf{p}=
\frac{\mu_{\alpha}^{*}}{\sqrt{\mu_{\alpha^{*}}^{2}+\mu_{\delta}^{2}}}
\left(
      \begin{array}{c}
      -\cos\alpha\sin\delta \\
       -\sin\alpha\sin\delta\\
       \cos\delta\\
       \end{array}
\right)
-
\frac{\mu_{\delta}}{\sqrt{\mu_{\alpha^{*}}^{2}+\mu_{\delta}^{2}}}
\left(
      \begin{array}{c}
      -\sin\alpha \\
       \cos\alpha\\
       0\\
       \end{array}
\right),
\end{equation}
where $\mu_{\alpha^{*}}=\mu_{\alpha}\cos\delta$. The determination of the celestial coordinates $(\alpha^{p},\delta^{p})$ of the pole requires transformation to an equatorial polar coordinate system. This yields
\begin{equation}\label{eq.A2}
\alpha^{p}=\arctan\left(\frac{-\mu_{\alpha}\cos\delta\sin\delta\sin\alpha-\mu_{\delta}\cos\alpha}{-\mu_{\alpha}\cos\delta\sin\delta\cos\alpha+\mu_{\delta}\sin\alpha}\right){
}\,
\end{equation}

\begin{equation}\label{eq.A3}
\delta^{p}=\arctan\left(\frac{\mu_{\alpha}\cos^{2}\delta}{\sqrt{(\mu_{\alpha}\cos\delta\sin\delta)^{2}+\mu_{\delta}^{2}}}\right){
}\,.
\end{equation}
The corresponding errors $(\sigma_{\alpha^{p}},\sigma_{\delta^{p}})$ are obtained by error propagation of Eqs.~(\ref{eq.A2}) and (\ref{eq.A3}). We have to first order

\begin{equation}
\sigma_{\alpha^{p}}^{2}=\left(\frac{\partial\alpha^{p}}{\partial\alpha}\right)^{2}\sigma_{\alpha}^{2}+\left(\frac{\partial\alpha^{p}}{\partial\delta}\right)^{2}\sigma_{\delta}^{2}+\left(\frac{\partial\alpha^{p}}{\partial\mu_{\alpha^{*}}}\right)^{2}\sigma_{\mu_{\alpha^{*}}}^{2}+\left(\frac{\partial\alpha^{p}}{\partial\mu_{\delta}}\right)^{2}\sigma_{\mu_{\delta}}^{2}
\end{equation}
and
\begin{equation}
\sigma_{\delta^{p}}^{2}=\left(\frac{\partial\delta^{p}}{\partial\alpha}\right)^{2}\sigma_{\alpha}^{2}+\left(\frac{\partial\delta^{p}}{\partial\delta}\right)^{2}\sigma_{\delta}^{2}+\left(\frac{\partial\delta^{p}}{\partial\mu_{\alpha^{*}}}\right)^{2}\sigma_{\mu_{\alpha^{*}}}^{2}+\left(\frac{\partial\delta^{p}}{\partial\mu_{\delta}}\right)^{2}\sigma_{\mu_{\delta}}^{2}.
\end{equation}
The partial derivatives are given as

\begin{equation}
\frac{\partial\alpha^{p}}{\partial\alpha}=\frac{1-\frac{(-\mu_{\alpha}^{*}\sin\delta\sin\alpha-\mu_{\delta}\cos\alpha)(\mu_{\alpha}^{*}\sin\delta\sin\alpha+\mu_{\delta}\cos\alpha)}{(-\mu_{\alpha}^{*}\sin\delta\cos\alpha+\mu_{\delta}\sin\alpha)^{2}}}{1+\left(\frac{-\mu_{\alpha}^{*}\sin\delta\sin\alpha-\mu_{\delta}\cos\alpha}{-\mu_{\alpha}^{*}\sin\delta\cos\alpha+\mu_{\delta}\sin\alpha}\right)^{2}},
\end{equation}

\begin{equation}
\frac{\partial\alpha^{p}}{\partial\delta}=-\frac{\frac{\mu_{\alpha}^{*}\cos\delta\sin\alpha}{-\mu_{\alpha}^{*}\sin\delta\cos\alpha+\mu_{\delta}\sin\alpha}}{1+\left(\frac{-\mu_{\alpha}^{*}\sin\delta\sin\alpha-\mu_{\delta}\cos\alpha}{-\mu_\alpha^{*}\sin\delta\cos\alpha+\mu_{\delta}\sin\alpha}\right)^{2}}+\frac{\frac{(-\mu_{\alpha}^{*}\sin\delta\sin\alpha-\mu_{\delta}\cos\alpha)\mu_{\alpha}^{*}\cos\delta\cos\alpha}{(-\mu_{\alpha}^{*}\sin\delta\cos\alpha+\mu_{\delta}\sin\alpha)^{2}}}{1+\left(\frac{-\mu_{\alpha}^{*}\sin\delta\sin\alpha-\mu_{\delta}\cos\alpha}{-\mu_\alpha^{*}\sin\delta\cos\alpha+\mu_{\delta}\sin\alpha}\right)^{2}},
\end{equation}

\begin{equation}
\frac{\partial\alpha^{p}}{\partial\mu_{\alpha}^{*}}=-\frac{\frac{\sin\delta\sin\alpha}{-\mu_{\alpha}^{*}\sin\delta\cos\alpha+\mu_{\delta}\sin\alpha}}{1+\left(\frac{-\mu_{\alpha}^{*}\sin\delta\sin\alpha-\mu_{\delta}\cos\alpha}{-\mu_\alpha^{*}\sin\delta\cos\alpha+\mu_{\delta}\sin\alpha}\right)^{2}}\hspace{0.1cm}+\hspace{0.1cm}\frac{\frac{(-\mu_{\alpha}^{*}\sin\delta\sin\alpha-\mu_{\delta}\cos\alpha)\sin\delta\cos\alpha}{(-\mu_{\alpha}^{*}\sin\delta\cos\alpha+\mu_{\delta}\sin\alpha)^{2}}}{1+\left(\frac{-\mu_{\alpha}^{*}\sin\delta\sin\alpha-\mu_{\delta}\cos\alpha}{-\mu_\alpha^{*}\sin\delta\cos\alpha+\mu_{\delta}\sin\alpha}\right)^{2}},
\end{equation}

\begin{equation}
\frac{\partial\alpha^{p}}{\partial\mu_{\delta}}=-\frac{\frac{\cos\alpha}{-\mu_{\alpha}^{*}\sin\delta\cos\alpha+\mu_{\delta}\sin\alpha}}{1+\left(\frac{-\mu_{\alpha}^{*}\sin\delta\sin\alpha-\mu_{\delta}\cos\alpha}{-\mu_\alpha^{*}\sin\delta\cos\alpha+\mu_{\delta}\sin\alpha}\right)^{2}}\hspace{0.2cm}-\hspace{0.2cm}\frac{\frac{(-\mu_{\alpha}^{*}\sin\delta\sin\alpha-\mu_{\delta}\cos\alpha)\sin\alpha}{(-\mu_{\alpha}^{2}\sin\delta\cos\alpha+\mu_{\delta}\sin\alpha)^{2}}}{1+\left(\frac{-\mu_{\alpha}^{*}\sin\delta\sin\alpha-\mu_{\delta}\cos\alpha}{-\mu_\alpha^{*}\sin\delta\cos\alpha+\mu_{\delta}\sin\alpha}\right)^{2}},
\end{equation}

\begin{equation}
\frac{\partial\delta^{p}}{\partial\alpha}=0,
\end{equation}

\begin{equation}
\frac{\partial\delta^{p}}{\partial\delta}=\frac{-\frac{\mu_{\alpha}^{*}\sin\delta}{\sqrt{(\mu_{\alpha}^{*}\sin\delta)^{2}+\mu_{\delta}^{2}}}-\frac{(\mu_{\alpha}^{*}\cos\delta)^{2}\mu_{\alpha}^{*}\sin\delta}{((\mu_{\alpha}^{*}\sin\delta)^{2}+\mu_{\delta}^{2})^{3/2}}}{1+\frac{(\mu_{\alpha}^{*}\cos\delta)^{2}}{(\mu_{\alpha}^{*}\sin\delta)^{2}+\mu_{\delta}^{2}}},
\end{equation}

\begin{equation}
\frac{\partial\delta^{p}}{\partial\mu_{\alpha}^{*}}=\frac{\frac{\cos\delta}{\sqrt{(\mu_{\alpha}^{*}\sin\delta)^{2}+\mu_{\delta}^{2}}}-\frac{(\mu_{\alpha}^{*}\sin\delta)^{2}\cos\delta}{((\mu_{\alpha}^{*}\sin\delta)^{2}+\mu_{\delta}^{2})^{3/2}}}{1+\frac{(\mu_{\alpha}^{*}\cos\delta)^{2}}{(\mu_{\alpha}^{*}\sin\delta)^{2}+\mu_{\delta}^{2}}},
\end{equation}

\begin{equation}
\frac{\partial\delta^{p}}{\partial\mu_{\delta}}=\frac{-\mu_{\alpha}^{*}\mu_{\delta}\cos\delta}{((\mu_{\alpha}^{*}\sin\delta)^{2}+\mu_{\delta}^{2})^{3/2}\left(1+\frac{(\mu_{\alpha}^{*}\cos\delta)^{2}}{(\mu_{\alpha}^{*}\sin\delta)^{2}+\mu_{\delta}^{2}}\right)}.
\end{equation}

\section{Orthogonality error }

Let us assume that $\mathbf{\hat{e}_{cp}}$ is a unit vector that defines the direction of the CP $(\alpha_{cp},\delta_{cp})$ in the celestial sphere, and $\mathbf{p}$, the normalized polar vector as given by Eq.~(\ref{eq.A1}), defines the pole $(\alpha^{p},\delta^{p})$ of the great circle for a given star in  the group. We define $\vec{\kappa}=\kappa\,\mathbf{\hat{e}_{cp}}$ where $\kappa$ is the orthogonality error and denotes the amount of $\mathbf{p}$ that is projected in the $\mathbf{\hat{e}_{cp}}$ direction. It is written as
\begin{equation}
\vec{\kappa}=(\mathbf{p}\cdot\mathbf{\hat{e}_{cp}})\,\mathbf{\hat{e}_{cp}}=\underbrace{(\cos\theta)}_{\kappa}\,\mathbf{\hat{e}_{cp}},
\end{equation} 
where $\theta$ is the angle between $\mathbf{p}$ and $\mathbf{\hat{e}_{cp}}$. However, $\theta$ is also the angle between the planes that contain $\mathbf{p}$ and $\mathbf{\hat{e}_{cp}}$, which defines it as a spherical angle. It is given by the cosine formula in the spherical triangle that contains the CP  $(\alpha_{cp},\delta_{cp})$ and the pole $(\alpha^{p},\delta^{p})$ as
\begin{equation}
\kappa\equiv cos\theta=\sin\delta_{cp}\sin\delta^{p}+\cos\delta_{cp}\cos\delta^{p}\cos(\alpha_{cp}-\alpha^{p}).
\end{equation}
Whenever perfect parallelism of the stellar motions is achieved, $\mathbf{p}$ and $\mathbf{\hat{e}_{cp}}$ are orthogonal and $\kappa=0$. In practice, one needs to minimize $\kappa$ to search for the CP and moving group members. The corresponding error of $\kappa$ is given by error propagation and takes position and proper motion errors into account. It is given as
\begin{equation}
\begin{array}{l}
\sigma_{\kappa}^{2}=\left[-\cos\delta_{cp}\cos\delta^{p}\sin(\alpha^{p}-\alpha_{cp})\right]^{2}\sigma_{\alpha^{p}}^{2}
+\left[\sin\delta_{cp}\cos\delta^{p}-\cos\delta_{cp}\sin\delta^{p}\cos(\alpha^{p}-\alpha_{cp})\right]^{2}\sigma_{\delta^{p}}^{2}.
\end{array}
\end{equation}

\section{Error propagation}

The \rm{\sc{Hipparcos}} catalog provides the five astrometric parameters $(\alpha,\delta,\mu_{\alpha}^{*},\mu_{\delta},\pi)$ together with the \textit{full} covariance matrix. Thus, the propagation of errors must consider the covariances between all observables. The complete expression for the error propagation is given as

\begin{eqnarray}\label{eq.c1}
\sigma_{\alpha^{p}}^{2}=\left(\frac{\partial\alpha^{p}}{\partial\alpha}\right)^{2}\sigma_{\alpha}^{2}+\left(\frac{\partial\alpha^{p}}{\partial\delta}\right)^{2}\sigma_{\delta}^{2}+\left(\frac{\partial\alpha^{p}}{\partial\mu_{\alpha}^{*}}\right)^{2}\sigma_{\mu_{\alpha^{*}}}^{2}+\left(\frac{\partial\alpha^{p}}{\partial\mu_{\delta}}\right)^{2}\sigma_{\mu_{\delta}}^{2}+2\left(\frac{\partial\alpha^{p}}{\partial\alpha}\right)\left(\frac{\partial\alpha^{p}}{\partial\delta}\right)\rho_{\alpha\delta}\,\sigma_{\alpha}\sigma_{\delta}+2\left(\frac{\partial\alpha^{p}}{\partial\alpha}\right)\left(\frac{\partial\alpha^{p}}{\partial\mu_{\alpha}^{*}}\right)\rho_{\alpha\mu_{\alpha}^{*}}\,\sigma_{\alpha}\sigma_{\mu_{\alpha}^{*}} + \nonumber \\ +2\left(\frac{\partial\alpha^{p}}{\partial\alpha}\right)\left(\frac{\partial\alpha^{p}}{\partial\mu_{\delta}}\right)\rho_{\alpha\mu_{\delta}}\,\sigma_{\alpha}\sigma_{\mu_{\delta}}+2\left(\frac{\partial\alpha^{p}}{\partial\delta}\right)\left(\frac{\partial\alpha^{p}}{\partial\mu_{\alpha}^{*}}\right)\rho_{\delta\mu_{\alpha}^{*}}\,\sigma_{\delta}\sigma_{\mu_{\alpha}^{*}}+2\left(\frac{\partial\alpha^{p}}{\partial\delta}\right)\left(\frac{\partial\alpha^{p}}{\partial\mu_{\delta}}\right)\rho_{\delta\mu_{\delta}}\,\sigma_{\delta}\sigma_{\mu_{\delta}}+2\left(\frac{\partial\alpha^{p}}{\partial\mu_{\alpha}^{*}}\right)\left(\frac{\partial\alpha^{p}}{\partial\mu_{\delta}}\right)\rho_{\mu_{\alpha}^{*}\mu_{\delta}}\,\sigma_{\mu_{\alpha}^{*}}\sigma_{\mu_{\delta}}
\end{eqnarray}
and
\begin{eqnarray}\label{eq.c2}
\sigma_{\delta^{p}}^{2}=\left(\frac{\partial\delta^{p}}{\partial\alpha}\right)^{2}\sigma_{\alpha}^{2}+\left(\frac{\partial\delta^{p}}{\partial\delta}\right)^{2}\sigma_{\delta}^{2}+\left(\frac{\partial\delta^{p}}{\partial\mu_{\alpha}^{*}}\right)^{2}\sigma_{\mu_{\alpha^{*}}}^{2}+\left(\frac{\partial\delta^{p}}{\partial\mu_{\delta}}\right)^{2}\sigma_{\mu_{\delta}}^{2}+2\left(\frac{\partial\delta^{p}}{\partial\alpha}\right)\left(\frac{\partial\delta^{p}}{\partial\delta}\right)\rho_{\alpha\delta}\,\sigma_{\alpha}\sigma_{\delta}+2\left(\frac{\partial\delta^{p}}{\partial\alpha}\right)\left(\frac{\partial\delta^{p}}{\partial\mu_{\alpha}^{*}}\right)\rho_{\alpha\mu_{\alpha}^{*}}\,\sigma_{\alpha}\sigma_{\mu_{\alpha}^{*}} + \nonumber \\ +2\left(\frac{\partial\delta^{p}}{\partial\alpha}\right)\left(\frac{\partial\delta^{p}}{\partial\mu_{\delta}}\right)\rho_{\alpha\mu_{\delta}}\,\sigma_{\alpha}\sigma_{\mu_{\delta}}+2\left(\frac{\partial\delta^{p}}{\partial\delta}\right)\left(\frac{\partial\delta^{p}}{\partial\mu_{\alpha}^{*}}\right)\rho_{\delta\mu_{\alpha}^{*}}\,\sigma_{\delta}\sigma_{\mu_{\alpha}^{*}}+2\left(\frac{\partial\delta^{p}}{\partial\delta}\right)\left(\frac{\partial\delta^{p}}{\partial\mu_{\delta}}\right)\rho_{\delta\mu_{\delta}}\,\sigma_{\delta}\sigma_{\mu_{\delta}}+2\left(\frac{\partial\delta^{p}}{\partial\mu_{\alpha}^{*}}\right)\left(\frac{\partial\delta^{p}}{\partial\mu_{\delta}}\right)\rho_{\mu_{\alpha}^{*}\mu_{\delta}}\,\sigma_{\mu_{\alpha}^{*}}\sigma_{\mu_{\delta}},
\end{eqnarray}
where $\rho_{xy}$ denotes the correlation coefficient of the observables $x$ and $y$. The partial derivatives are given in Appendix~A. 

\end{document}